\RequirePackage{fix-cm}
\documentclass[smallextended]{svjour3}       
\smartqed  
\usepackage{graphicx}
\usepackage{subfigure}
\usepackage{color}
\usepackage{natbib} 
\usepackage{bm}
\usepackage{epstopdf}
\usepackage{here}
\usepackage{mathptmx}      
\usepackage{amssymb}
\newcommand{\argmin}{\mathop{\rm arg~min}\limits}
\begin{document}
\title{Spatiotemporal Superresolution Measurement based on POD and Sparse Regression applied to a Supersonic Jet measured by PIV and Near-field Microphone}

\titlerunning{Spatio-temporal Superresolution Measurement}        

\author{Yuta Ozawa, Takayuki Nagata, \\and Taku Nonomura}

\authorrunning{Y. Ozawa et al.} 

\institute{Y. Ozawa \at
              Aoba 6-6-01, Aramaki, Aoba-ku, Sendai, Miyagi, Japan
              Tel.: +81-22-795-4075\\
              Fax: +81-22-678910\\
              \email{yuta.ozawa@tohoku.ac.jp}           
}

\date{Received: date / Accepted: date}
\maketitle

\begin{abstract}
The present study proposed the framework of the spatiotemporal superresolution measurement based on the and sparse regression with dimensionality reduction using the proper orthogonal decomposition (POD). The non-time-resolved particle image velocimetry (PIV) and the time-resolved near-field acoustic measurements using microphones were simultaneously performed for a Mach 1.35 supersonic jet. POD is applied to PIV and microphone data matrices and the sparse linear regression model of the reduced-order data is calculated using the least absolute shrinkage and selection operator regression. The effects of the hyperparameters of the superresolution measurement were quantitatively evaluated through randomized cross-validation. The superresolved velocity field indicated the smooth convection of the velocity fluctuations associated with the screech tone, while the convection of the large-scale structures at the downstream side was not observed. The proposed framework can reconstruct the unsteady fluctuation with multiple frequency phenomena, although the reconstruction is limited to the phenomena that is associated with the microphone output.
\end{abstract}

\section*{Nomenclature}

\noindent{Scalar variables} \\
\vbox{\noindent\setlength{\tabcolsep}{0mm}%
\begin{tabular}{p{25mm}cl} %
\hfil$D$       \hfil & :\hspace{4mm} & nozzle exit diameter \\
\hfil$E$       \hfil & :\hspace{4mm} & reconstruction error \\
\hfil$f$       \hfil & :\hspace{4mm} & frequency \\
\hfil$f_s$     \hfil & :\hspace{4mm} & peak frequency of screcch tone \\
\hfil$L_{sh}$  \hfil & :\hspace{4mm} & shock cell length \\
\hfil$m$       \hfil & :\hspace{4mm} & number of data points in space \\
\hfil$M$       \hfil & :\hspace{4mm} & Mach number \\
\hfil$n$       \hfil & :\hspace{4mm} & number of data points in time \\
\hfil$n_{td}$  \hfil & :\hspace{4mm} & number of data points for time-delay \\
\hfil$N$       \hfil & :\hspace{4mm} & total number of PIV snapshots \\
\hfil$p_{j}$   \hfil & :\hspace{4mm} & microphone data acquired by $j$th microphone \\
\hfil$PR$      \hfil & :\hspace{4mm} & projection ratio \\
\hfil$r$       \hfil & :\hspace{4mm} & number of POD modes for the dimensionality reduction \\
\hfil$St$      \hfil & :\hspace{4mm} & Strouhal number \\
\hfil$t_{i}$   \hfil & :\hspace{4mm} & discrete-time \\
\hfil$u,v$     \hfil & :\hspace{4mm} & streamwise and radial velocity \\
\hfil$u_c$     \hfil & :\hspace{4mm} & convection velocity \\
\hfil$U_{j}$   \hfil & :\hspace{4mm} & streamwise velocity at the nozzle exit derived under \\
\hfil          \hfil &  \hspace{4mm} & the assumption of the isentropic flow \\
\hfil$\kappa$  \hfil & :\hspace{4mm} & sampling-rate ratio of the acoustic and PIV measurements \\
\hfil$\lambda$ \hfil & :\hspace{4mm} & regularization parameter for group-LASSO regression \\
\end{tabular}}

\noindent{Vectors and Matrices} \\
\vbox{\noindent\setlength{\tabcolsep}{0mm}%
\begin{tabular}{p{25mm}cl} %
\hfil$\mathbf{M}$  \hfil & :\hspace{4mm} & time-delay embedded microphone data matrix \\
\hfil$\mathbf{S}$  \hfil & :\hspace{4mm} & diagonal matrix of singular values \\
\hfil$\mathbf{u}$  \hfil & :\hspace{4mm} & column vector of streamwise velocity components \\
\hfil$\mathbf{U}$  \hfil & :\hspace{4mm} & orthogonal spatial modes matrix\\
\hfil$\mathbf{v}$  \hfil & :\hspace{4mm} & column vector of radial velocity components \\
\hfil$\mathbf{V}$  \hfil & :\hspace{4mm} & orthogonal temporal modes matrix\\
\hfil$\mathbf{X}$  \hfil & :\hspace{4mm} & Data matrix\\
\hfil$\mathbf{Z}$  \hfil & :\hspace{4mm} & POD coefficients matrix\\
\hfil$\mathbf{Z_{\mathrm{\Phi}}}$  \hfil & :\hspace{4mm} & projection of
the $\mathbf{Z_{\mathrm{PIV}}}$ onto the $\mathrm{\Phi}$ space \\
\hfil$\mathrm{\Psi}$  \hfil & :\hspace{4mm} & regression coefficients matrix \\
\hfil$\mathrm{\Phi}$  \hfil & :\hspace{4mm} & regression coefficients matrix \\
\hfil                 \hfil &  \hspace{4mm} & consists of the non-zero components of $\mathrm{\Psi}$ \\
\end{tabular}}

\noindent{Superscripts} \\
\vbox{\noindent\setlength{\tabcolsep}{0mm}%
\begin{tabular}{p{25mm}cl}
\hfil$\bar{}$        \hfil & :\hspace{4mm} & time-averaged component \\
\hfil$\widetilde{}$  \hfil & :\hspace{4mm} & fluctuation component \\
\hfil$\hat{}$        \hfil & :\hspace{4mm} & estimated component \\
\hfil$'$             \hfil & :\hspace{4mm} & downsampled component \\
\end{tabular}}

\noindent{Subscripts} \\
\vbox{\noindent\setlength{\tabcolsep}{0mm}%
\begin{tabular}{p{25mm}cl}
\hfil$\mathrm{MIC}$      \hfil & :\hspace{4mm} & microphone data \\
\hfil$\mathrm{PIV}$      \hfil & :\hspace{4mm} & PIV data \\
\end{tabular}}

\section{Introduction}
The exhaust flow of a supersonic aircraft engine emits strong acoustic waves and causes serious noise pollution around the airport. Therefore, a lot of studies have been performed for investigating supersonic jet noise in past decades. When the jet flow contains shock waves, the peaky noise called the screech tone dominates the acoustic field. \citet{powell1953mechanism,powell1953noise} firstly proposed the generation mechanism of the screech tone and a lot of studies have been dedicated to this field for a long time \citep{bailly2016high,raman1999supersonic,tam1995supersonic}. The screech tone is generated due to the acoustic feedback loop that consists of turbulent structures and acoustic waves. The turbulent structures at the nozzle lip develop with their convection, and the interference of the shock waves and the turbulent structures generates the acoustic waves. Then, the acoustic waves propagating upstream excite the instability waves at the nozzle lip. The interaction of shock waves and vortex structures was computationally investigated by \citet{suzuki2003shock} and they showed that the shock waves tend to leak near the saddle point of the vortex structures resulting in the acoustic radiation. This shock leakage is also observed by \citet{shariff2013ray}. For an axisymmetric cold jet in the screeching condition, there are four kinds of instability modes that are experimentally identified by the early study of \citet{powell1953mechanism}. The axisymmetric modes A1 and A2 dominate the aeroacoustic field when the Mach number is relatively low $(M_j\leq1.3)$, and the flapping mode B and the helical mode C appear at the higher Mach numbers. In addition to those modes, sinuous/flapping mode D was identified. The characteristics of these modes have been extensively investigated using both the computations and experiments \citep{panda1999experimental,andre2011experimental,edgington2014coherent,mercier2016schlieren}.

Although the physical mechanism and characteristics of the screech tone are well understood based on the statistical data, the time-resolved data of the unsteady dynamics is limited. Numerical simulations are often performed for the discussion of the unsteady dynamics of the screech tone \citep{gojon2017numerical,arroyo2019identification,li2020acoustic}. However, a large-eddy simulation that can compute the aeroacoustic field of the high-Reynolds-number jet requires excessive computational cost and is not suitable for parametric analysis, see e.g. \cite{nonomura2019large,nonomura2021computational,pineau2021numerical,pineau2021links}. On the other hand, the experimental techniques are productive while the measurable quantities are limited. The recent technology advancement allows experimentally measuring the time-resolved data using a high-speed camera, and the data-driven analysis such as the proper orthogonal decomposition (POD) \citep{berkooz1993proper} or dynamic mode decomposition \citep{schmid2010dynamic} is often employed. There are many attempts to identify the screech tone from the time-resolved schlieren images \citep{li2021screech,lim2020short,rao2020screech,ozawa2018identification,mercier2017experimental,ohmizu2022demonstration}. However, a conventional high-speed camera still does not have a sufficient sampling rate for the visualization of the entire flow field with high spatial resolution. As an example, a Phantom V2640 camera, which is a fastest high-speed camera with a 4M pixels image sensor, has the maximum sampling rate of 6.6 kHz at full pixels ($2048\times1920$ pixels). Considering a laboratory-scale supersonic jet with a diameter of 10~mm and a velocity of 400~m/s, the shear layer thickness is 1~mm or less. When the spatial resolution is assumed to be 0.1~mm/pixels that is 1/10 of the shear layer thickness, the sampling rate of 500~kHz (2~$\mu$s) can achieve the displacement on the image of 8~pixels that can visualize the unsteady dynamics. Therefore, there are no high-speed cameras that can visualize the supersonic flow with a sufficiently high sampling rate with maintaining the full image resolution, though the corresponding PIV spatial resolution is acceptable but not significantly high due to its interrogation window. Here, it should be noted that although the spatial resolution substantially decreases due to the interrogation window when applying the image correlation such as the particle image velocimetry (PIV) or background oriented schlieren (BOS) \citep{lee2021comparison,tan2019correlation,edgington2018upstream}, the improvement of the spatial resolution for the image correlation is out of scope in the present study while there are some studies regarding this topic. For example, one of the methods for the improvement of spatial resolution is the single-pixel ensemble correlation method, but its application is limited to the time-averaged velocity field \citep{westerweel2004single,scharnowski2012reynolds,ozawa2020single,nonomura2021generalized}. Therefore, the improvement of spatial resolution for the instantaneous fields is not straightforward and is left for future work. Additionally, time-resolved PIV for a high-speed flow requires an expensive high-repetition burst laser system\citep{price2021supersonic,beresh2015pulse}. Owing to the insufficient camera performance or the expensive burst laser system described above, the spatiotemporal superresolution measurement, which is a reconstruction of the time-resolved data from the non-time-resolved data acquired with the existing experimental devices, can be a powerful experimental technique for analyzing high-speed flow phenomena.

The reconstruction of the time-resolved data from the non-time-resolved data has been performed in various flow fields using stochastic estimation and sensor fusion\citep{nickels2020low,zhang2020spectral,li2021pressure}. Those works rely on the linear stochastic estimation (LSE) combined with POD and the low-dimensional dynamics of the energy-containing structures in high-Reynolds-number flow was estimated. \citet{tinney2008low} performed PIV and near field pressure measurements of a $M_j=0.85$ axisymmetric jet and constructed a reduced-order model based on the Fourier-azimuthal decomposition and POD. \citet{tu2013integration} performed the time-resolved PIV and hot-wire measurements of a wake flow behind a model and estimated the time-resolved velocity field using the downsampled PIV data and time-resolved hot-wire data. The reconstruction of the time-resolved velocity field is based on the multi-time-delay modified linear stochastic estimation (MTD-mLSE) \citep{durgesh2010multi} that can strengthen the correlation of each coefficient. The proposed method effectively reconstructs the unsteady dynamics of the wake structures. It should be noted that the spatiotemporal superresolution technique has strong relation with the low-dimensional modeling of flow fields and their applications \citep{brunton2019data,suzuki2020few,nankai2019linear,nonomura2021quantitative,kanda2021feasibility}. 

The present study applied this method to the aeroacoustic fields of an $M_j=1.35$ axisymmetric jet and reconstructed the time-resolved velocity fields. The ideally expanded supersonic jet was measured by means of the non-time-resolved PIV measurements and time-resolved near-field acoustic measurement using microphones. Thus far, there were some studies reconstructing the time-resolved velocity fields using pressure data \citep{li2021pressure,tinney2008low}, and the present authors also tried the reconstruction using the leading POD modes and the linear least square regression, as reported in \citet{ozawa2021pod}. However, the error was 99\% and its accuracy was disappointing, where the definition of the error is the same as that in the present study. This might be because of weak correlation between the microphone signals and the PIV modes. The present study extracts only the PIV modes associated with microphone signals using the least absolute shrinkage and selection operator (LASSO) regression \citep{tibshirani1996regression} and improves the estimation accuracy of the linear regression model. The quantitative evaluation of the proposed method was conducted based on the cross-validation and the effectiveness and limitation of the POD-based superresolution measurement are provided.

\section{Experimental setup}
A supersonic jet generating system installed in an anechoic room at Tohoku University was employed for the present experiment. Refer to \citet{ozawa2020effect,ozawa2020single} for more details on the experimental facilities. The underexpanded supersonic jet with $M_j=1.35$ was reproduced using a contoured convergent nozzle. The contour of the convergent nozzle was designed based on the reference~\citep{andre2013broadband}, and the diameter at the nozzle exit $D$ was set to be 10~mm. The nozzle pressure ratio was 2.97, and the Reynolds number based on the nozzle exit diameter was $4.62 \times 10^5$. The stagnation temperature was 297~K.

\begin{figure}[H]
    \centering
    \includegraphics[scale=0.8]{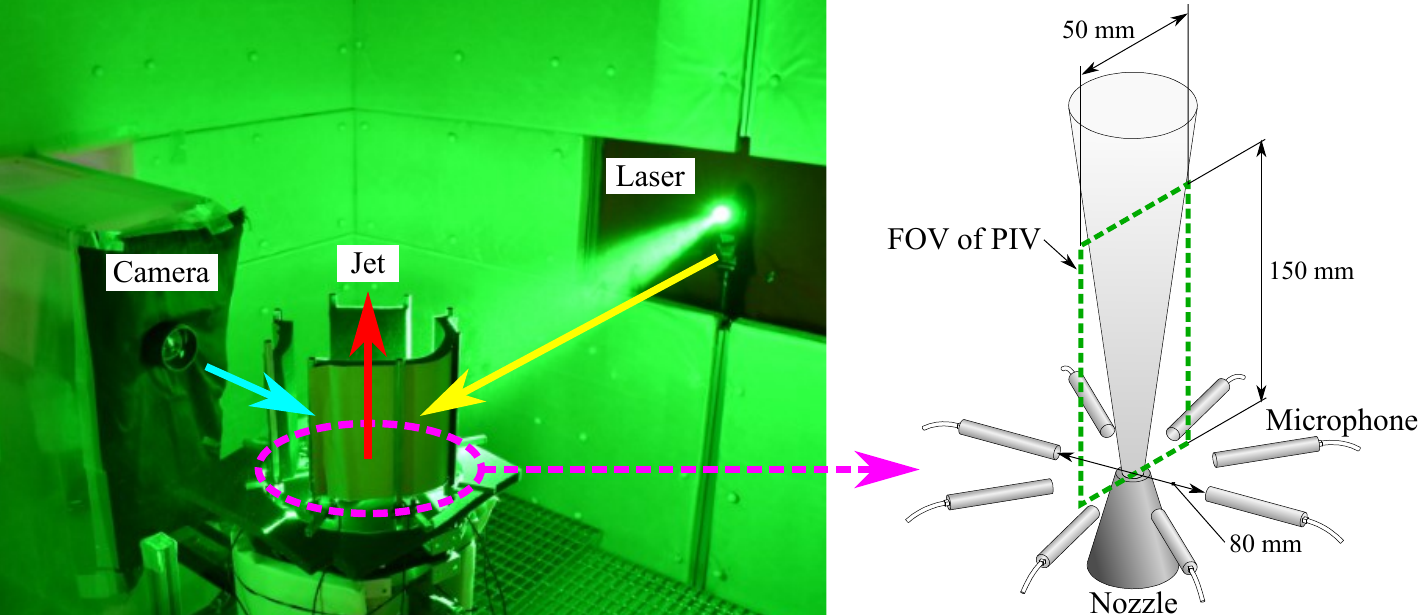}
    \caption{Schematic image of the PIV plane and the microphone positions.}
    \label{fig:setup}
\end{figure}

The present study simultaneously performed the PIV and the near-field acoustic measurement using microphones. Figure~\ref{fig:setup} shows the schematic image of the experimental setup. The nozzle and the stagnation chamber are located in the center of the anechoic room and the jet flow towards the upper side. Eight microphones (TYPE4158N, ACO) were placed around the nozzle lip with keeping the distance of $r/D=4$, and the near-field acoustic measurement was conducted. The support for the microphones was covered by the sound-absorbing panel to prevent acoustic reflections. The acoustic signals were recorded using an amplifier (TYPE5006/4, ACO) and a data acquisition system (USB-6366, National Instruments). 

The high-speed camera (Phantom V1840, Vision Research) and the double-pulsed laser (LDY-300PIV, Litron) are installed in the anechoic room and the non-time-resolved planar PIV data are acquired. The field of view (FOV) of the PIV is $150 \times 50$~mm as shown in Fig.~\ref{fig:setup}. The camera lens (Nikkor 80--200~mm f/2.8, Nikon) and the 12~mm long extension tube were employed for the optics and the spatial resolution of the high-speed camera was set to be $2048\times776$ pixels. Tracer particles for PIV measurement are generated using a glycerin 50\% aqueous solution and Raskin nozzles. Raskin nozzles are incorporated into the jet generating system as well as the anechoic room. Refer to \citet{ozawa2020effect} for details. Therefore, both jet and ambient flow was fully seeded by the tracer particles. The diameter of the tracer particles is approximately several micrometers \citep{kahler2002generation}.

The PIV and acoustic measurements were synchronized using a trigger signal generated from the function generator (WF1974, NF). Table~\ref{tab:measurementparameter} summarizes the spatial and temporal resolution of each measurement. In contrast to the low sampling rate of the PIV system, that of the acoustic measurement is sufficiently high to resolve the dynamics of the large-scale structures. Here, the ratio of the sampling rate of the PIV and acoustic measurements was defined as $\kappa=50$. 

\begin{table}[H]
    \begin{center}
        \caption{Parameters of the measurements.}
            \hspace{1.6cm}
            \begin{tabular}{c c c} \hline
                & PIV & Acoustic measurement\\ \hline
                Spatial resolution & $256\times97$ vectors & 8 points\\
                Temporal resolution & 4,000 Hz & 200,000 Hz\\ 
                Number of dataset & 15,000 snapshots & 749,951 points\\ \hline
            \end{tabular}
        \label{tab:measurementparameter}
    \end{center}
\end{table}

\section{Calculation procedure of the superresolution}
\label{sec:CalculationProcedure}
\begin{figure}[H]
    \centering
    \includegraphics[scale=0.8]{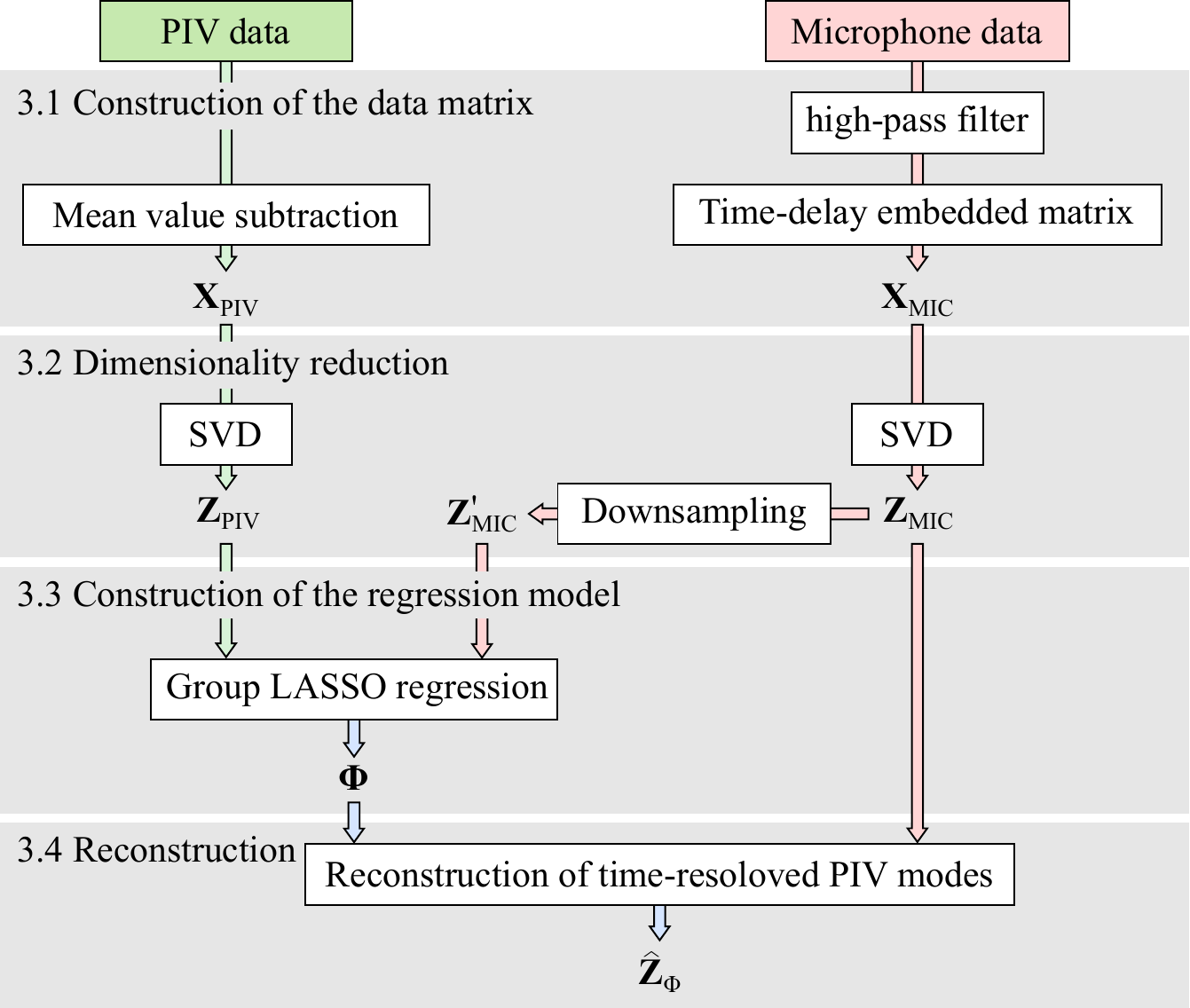}
    \caption{Flowchart of the data analysis for the superresolution.}
    \label{fig:flowchart}
\end{figure}

Figure~\ref{fig:flowchart} illustrates the flow chart of the data analysis for the proposed method. The first step of the superresolution is the construction of the data matrix using the simultaneously measured PIV and microphone data. Then, the reduced-order POD coefficients of each data matrix are obtained by applying the singular value decomposition (SVD) in the second step. The third step is the construction of the linear regression model that estimates the POD coefficients of the PIV from the microphone data. The calculated regression coefficients matrix is simply multiplied by the time-resolved POD coefficients matrix of the microphone data in the fourth step, and the time-resolved POD coefficients of the PIV data can be obtained. The detailed procedure of the superresolution is described in the following subsections.

\subsection{Definition of the data matrices}
\begin{figure}[H]
    \centering
    \includegraphics[scale=0.8]{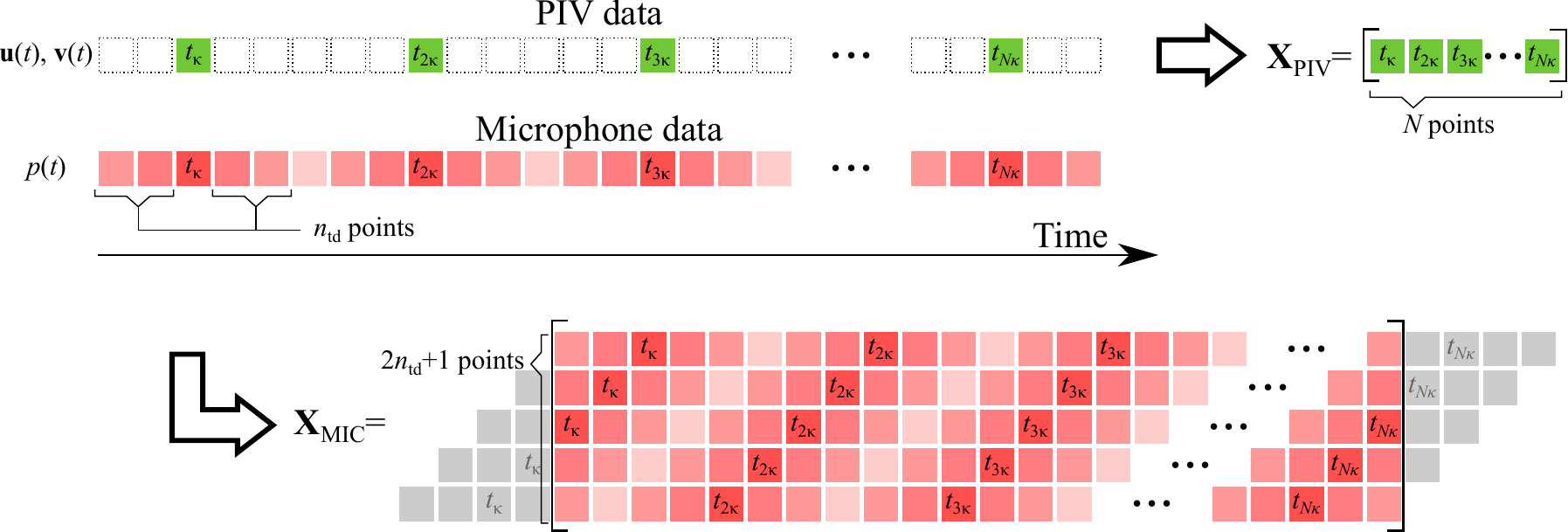}
    \caption{Schematic image of the data matrices definition.}
    \label{fig:matrixdef}
\end{figure}

The data matrices are firstly constructed using the acquired experimental data. Figure~\ref{fig:matrixdef} depicts the schematics of the data matrices definition. The colored boxes in this figure indicate the data point of each measurement. The data points of the PIV measurement are sparsely distributed because the sampling rate of the PIV is lower than that of the acoustic measurement. Note that this figure does not depict the actual sampling of the present experiment.

The PIV snapshots contain the two-dimensional components of the velocity field, $\mathbf{u}$ and $\mathbf{v}$, on the visualization plane. The present study subtracts the mean velocity from each snapshot and only the fluctuation components of the velocity are used:

\begin{equation}
\begin{array}{c}
u=\bar{u}+\widetilde{u}, \\
v=\bar{v}+\widetilde{v},
\end{array}
\label{eqn:u_decompose}
\end{equation}

\noindent
where notations $\bar{\circ}$ and $\widetilde{\circ}$ indicate the mean and fluctuation components, respectively. Here, the discrete-time $t_{i}$ $(1 \leq i \leq N\kappa)$ is defined based on the sampling rate of the acoustic measurement, where $N$ and $\kappa$ are the total number of PIV snapshots and the ratio of the sampling rate of the acoustic and PIV measurements, respectively. The PIV data matrix $\mathbf{X}_{\mathrm{PIV}}$ is constructed collecting the velocity fluctuation of each snapshot:

\begin{equation}
\mathbf{X}_{\mathrm{PIV}}=\left[\begin{array}{ccccc}
\widetilde{\mathbf{u}}(t_{\kappa}) & \widetilde{\mathbf{u}}(t_{2\kappa}) & \widetilde{\mathbf{u}}(t_{3\kappa}) & \cdots & \widetilde{\mathbf{u}}(t_{N\kappa}) \\
\widetilde{\mathbf{v}}(t_{\kappa}) & \widetilde{\mathbf{v}}(t_{2\kappa}) & \widetilde{\mathbf{v}}(t_{3\kappa}) & \cdots & \widetilde{\mathbf{v}}(t_{N\kappa}) \\
\end{array}\right],
\label{eqn:pivdata}
\end{equation}

\noindent
where $\widetilde{\mathbf{u}}(t_{i})$ and $\widetilde{\mathbf{v}}(t_{i})$ are the vector form of the streamwise and radial components of the measured velocity. The size of this matrix is $\mathbf{X}_{\mathrm{PIV}} \in \mathbb{R}^{2m \times N}$, where $m$ is the total number of velocity vectors.

The microphone data matrix $\mathbf{X}_{\mathrm{MIC}}$ is constructed using the acoustic signals. Microphone signals are first applied to the high-pass filter of 1~kHz and the noise due to the acoustic reflection is eliminated. Although the temporal resolution is sufficiently high to resolve the unsteady dynamics of the jet, the spatial resolution of the acoustic measurement is quite low compared with that of PIV, resulting in the less rank of the matrix. Moreover, the PIV and acoustic measurements acquire the different physical quantities that are not directly connected. Therefore, the reconstruction of the time-resolved velocity fields does not work well when the same formulation as the PIV data matrix is directly applied to the microphone data. To solve this issue, the time-delay embedded data matrix was constructed using the microphone data and the matrix rank increased. The present study defines that the time-delay $n_{\mathrm{td}}$ is the number of data points of both past and future at a given time as illustrated in Fig.~\ref{fig:matrixdef}. This is the same as the definition in the MTD-mLSE of the references \citep{durgesh2010multi,tu2013dynamic}. The time-delay embedded microphone data matrix is defined for each microphone as follows:

\begin{equation}
\mathbf{M}_{j}=\left[\begin{array}{ccccc}
p_{j}\left(t_{1-n_{\mathrm{td}}}\right) & p_{j}\left(t_{2-n_{\mathrm{td}}}\right) & p_{j}\left(t_{3-n_{\mathrm{td}}}\right) & \cdots &
p_{j}\left(t_{N\kappa-n_{\mathrm{td}}}\right) \\
\vdots & \vdots & \vdots & \ddots & \vdots \\
p_{j}\left(t_{1}\right) & p_{j}\left(t_{2}\right) & p_{j}\left(t_{3}\right) & \cdots & p_{j}\left(t_{N\kappa}\right) \\
\vdots & \vdots & \vdots & \ddots & \vdots \\
p_{j}\left(t_{1+n_{\mathrm{td}}}\right) & p_{j}\left(t_{2+n_{\mathrm{td}}}\right) & p_{j}\left(t_{3+n_{\mathrm{td}}}\right) & \cdots &
p_{j}\left(t_{N\kappa+n_{\mathrm{td}}}\right)
\end{array}\right],
\label{eqn:micdata_single}
\end{equation}

\noindent
where $p_{j}(t_{i})$ is the microphone data acquired by $j$th microphone ($1 \leq j \leq 8$). The size of this matrix is $\mathbf{M}_{j} \in \mathbb{R}^{(2n_{\mathrm{td}}+1)\times N\kappa}$. Finally, the matrix of each microphone was collected and the microphone data matrix $\mathbf{X}_{\mathrm{MIC}}$ is defined as follows:

\begin{equation}
 \mathbf{X}_{\mathrm{MIC}}=\left[\begin{array}{c}
\mathbf{M}_{1} \\
\mathbf{M}_{2} \\
\vdots \\
\mathbf{M}_{8}
\end{array}\right].
\label{eqn:micdata}
\end{equation}

\noindent
Here, time-delay $n_{\mathrm{td}}$ is one of the hyperparameters of the proposed method. The effect of $n_{\mathrm{td}}$ is discussed at Sec.~\ref{sec:taueffect}.

\subsection{Dimensionality reduction based on the POD}
\begin{figure}[H]
    \centering
    \includegraphics[scale=0.5]{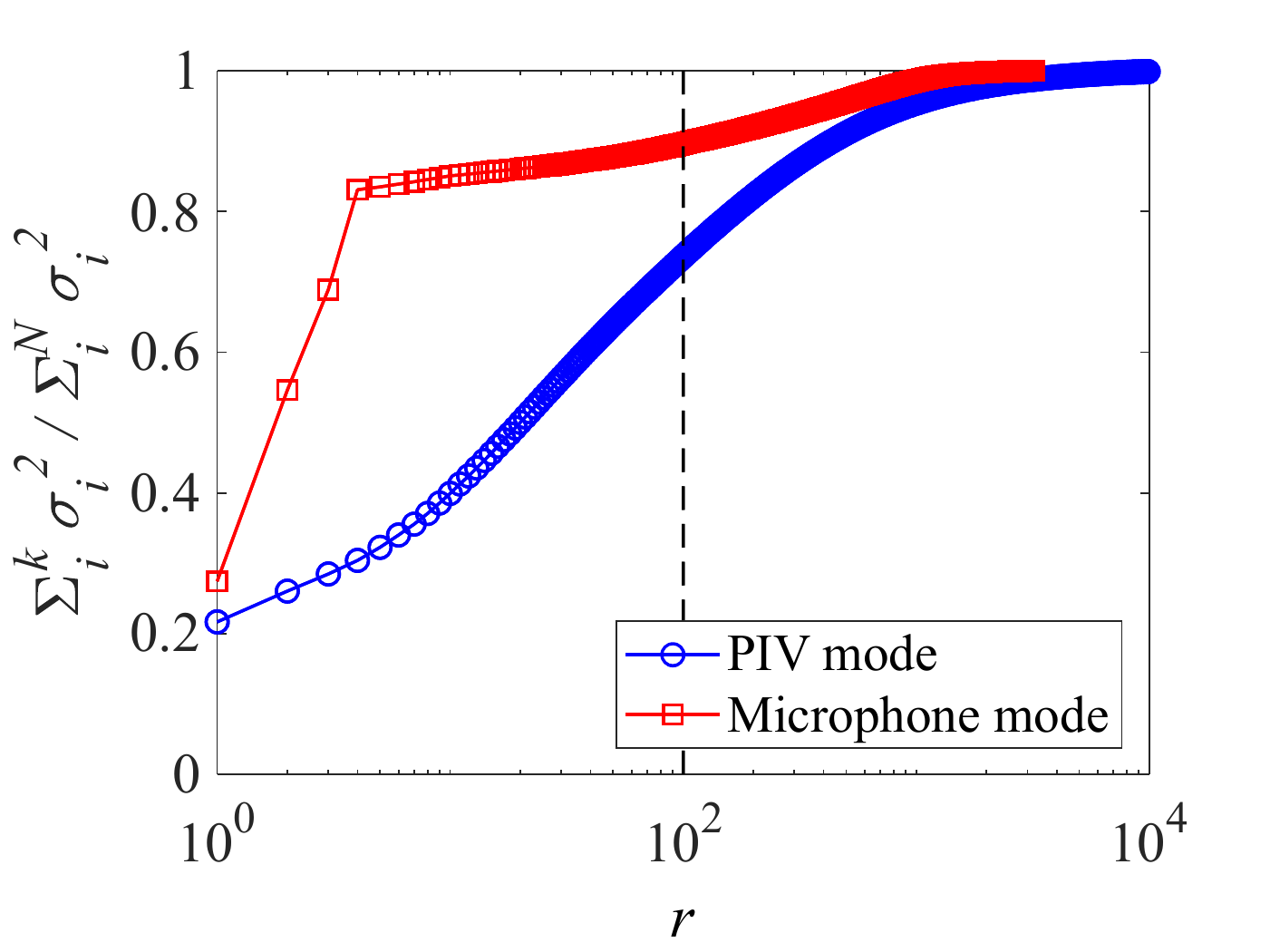}
    \caption{Partial amount of energy contained in the first $r$ POD modes.}
    \label{fig:POD-energy}
\end{figure}

The POD is a modal analysis method and extracts an orthogonal basis that expresses the data with the utmost efficiency \citep{berkooz1993proper,taira2017modal}. After \citet{lumley1967structure} firstly applied this analysis method into the fluid dynamics, the POD has been applied to various flow fields and extracted the coherent structures of turbulent flows. In application of the POD to fluid data, the data matrix $\mathbf{X} \in \mathbb{R}^{m \times n}$ is constructed using a column vector of fluid data at a given time. Here, $m$ and $n$ are the number of data points in space and time, respectively. The POD modes are simply calculated as the SVD as follows:

\begin{equation}
\mathbf{X}=\mathbf{U S V}^\mathsf{T},
\label{eqn:svd}
\end{equation}

\noindent
where $\mathbf{U} \in \mathbb{R}^{m \times m}$, $\mathbf{S} \in \mathbb{R}^{m \times n}$, and $\mathbf{V} \in \mathbb{R}^{n \times n}$ are orthogonal spatial modes, a matrix of which diagonal components are singular values $\sigma$, and orthogonal temporal modes. The singular values $\sigma$ represent the contribution of each mode with respect to the original data, and they are arranged in the matrix $\mathbf{S}$ in descending order in the present study. Therefore, equation~\ref{eqn:svd} can be interpreted that the spatial mode $\mathbf{U}$ evolves with POD modes coefficients $\mathbf{S V}^\mathsf{T}$. Here, the dimensionality reduced matrix of the original data matrix $\mathbf{X}^{(r)}$ is calculated using the first $r$ POD modes:

\begin{equation}
\mathbf{X}^{(r)}=\mathbf{U}^{(r)} \mathbf{Z},
\label{eqn:svd_reduction}
\end{equation}

\begin{equation}
\mathbf{Z}=\mathbf{S}^{(r)} \mathbf{V}^{(r)\mathsf{T}},
\label{eqn:podcoef}
\end{equation}

\noindent
where $\mathbf{Z}$ is the POD coefficients of the first $r$ modes. In the present study, the SVD was applied to the data matrices of the PIV $\mathbf{X}_\mathrm{PIV}$ and microphones $\mathbf{X}_\mathrm{MIC}$, and those dimensionality reduced matrices were obtained with the rank of $r_{\mathrm{PIV}}$ and $r_{\mathrm{MIC}}$, respectively:

\begin{equation}
\mathbf{Z}_\mathrm{PIV}=\mathbf{S}_\mathrm{PIV}^{(r_{\mathrm{PIV}})} \mathbf{V}_\mathrm{PIV}^{(r_{\mathrm{PIV}})\mathsf{T}},
\label{eqn:pivcof}
\end{equation}

\begin{equation}
\mathbf{Z}_\mathrm{MIC}=\mathbf{S}_\mathrm{MIC}^{(r_{\mathrm{MIC}})} \mathbf{V}_\mathrm{MIC}^{(r_{\mathrm{MIC}})\mathsf{T}},
\label{eqn:miccof}
\end{equation}

\noindent
where $\mathbf{Z}_{\mathrm{PIV}} \in \mathbb{R}^{r_{\mathrm{PIV}} \times N}$ and $\mathbf{Z}_{\mathrm{MIC}} \in \mathbb{R}^{r_{\mathrm{MIC}} \times N\kappa}$. Figure~\ref{fig:POD-energy} shows the partial amount of energy contained in the first $r$ POD modes. In this figure, the PIV modes were calculated using 15,000 snapshots, and the microphone modes were calculated from the microphone data matrix of $n_{\mathrm{td}}=200$. Since the supersonic jet is fully turbulent, the modal decomposition of PIV data is not efficient. On the other hand, the energy of the first four microphone modes is high. This is because the acoustic field is dominated by the screech tone and the microphone modes associated with the screech tone are efficiently extracted. Here, the present study employed the first 100 PIV and microphone modes for the superresolution measurement ($r_{\mathrm{PIV}}=r_{\mathrm{MIC}}=100$). The energy contained in the first 100 PIV and microphone modes was approximately 73$\%$ and 89$\%$, respectively. Therefore, even after the dimensionality reduction, the PIV and microphone modes still have sufficient energy to express the original aeroacoustic field. In the present study, the row vectors of $\mathbf{Z}_\mathrm{PIV}$ and $\mathbf{Z}_\mathrm{MIC}$ are referred to as the PIV and microphone modes, respectively. 

\subsection{Construction of the regression model}
\begin{figure}[H]
    \centering
    \includegraphics[scale=0.6]{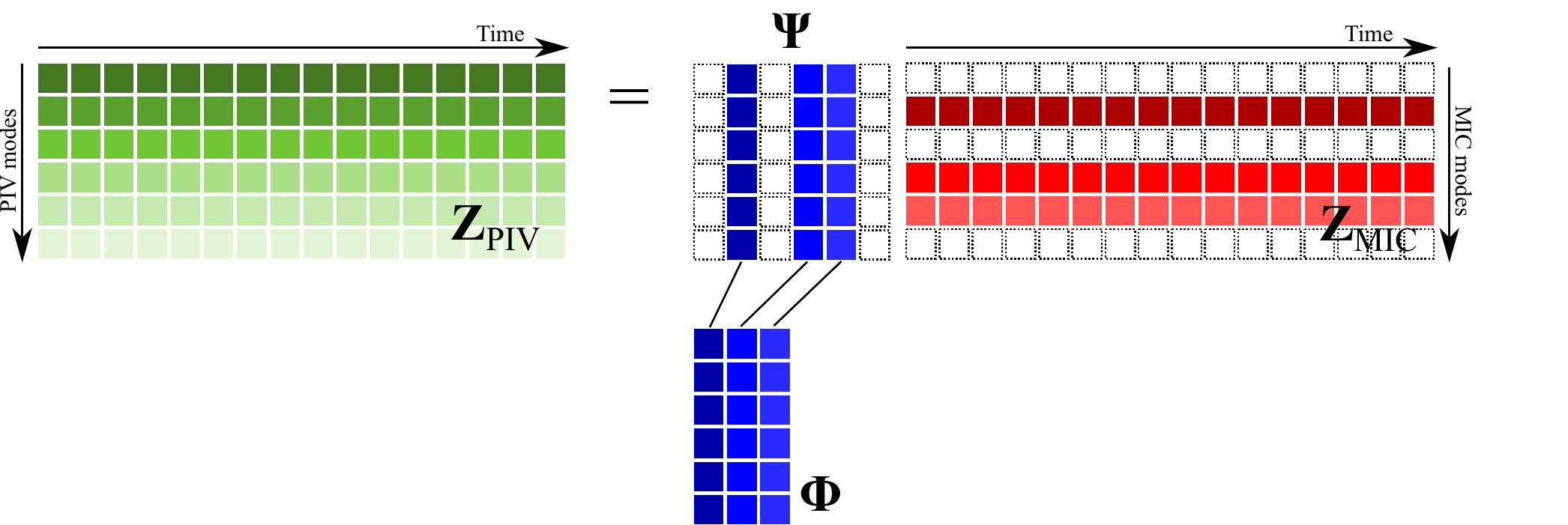}
    \caption{Schematic image of the regression model.}
    \label{fig:regressionimage}
\end{figure}

The present study construct the linear regression model using the POD coefficients of the PIV and microphone data matrices defined in the previous section. The PIV and microphone data of a screeching jet may have a strong correlation because the aeroacoustic field is dominated by the screech phenomena. Therefore, the present study firstly applied the linear regression for the PIV and microphone data of a supersonic jet and evaluate its applicability and limitations. The superresolution measurement assumes that the PIV modes can be correlated with the microphone modes as a form of a linear regression model:

\begin{equation}
\mathbf{Z}_\mathrm{PIV}=\mathrm{\Psi}\mathbf{Z}_\mathrm{MIC}^{\prime},
\label{eqn:regression}
\end{equation}

\noindent
where $\mathrm{\Psi}$ is the regression coefficients matrix, and $\mathbf{Z}_\mathrm{MIC}^{\prime}$ is a downsampled matrix of $\mathbf{Z}_\mathrm{MIC}$, which consists of the extracted row vectors of $\mathbf{Z}_\mathrm{MIC}$ at the timing when the PIV snapshots exist. In other words, the PIV modes can be estimated from only regression coefficients and the time-resolved microphone modes. Although the dimensionality reduction was applied to the microphone data, all of the phenomena in PIV modes cannot be fully estimated from the dominant microphone modes. This is because the generation of the acoustic waves measured by microphones is caused by a small part of the fluid fluctuation. Thus far, the regression of the largest microphone modes to the largest PIV modes was tried, but the flow feature was not recovered and the error of the reconstruction became no less than 98 \% \citep{nishikori2022superresolution}. Consequently, the present study improves the estimation accuracy of the linear regression model using only the microphone modes of which correlation with the PIV modes is high. This is realized using the LASSO regression \citep{tibshirani1996regression} that is the regression analysis method incorporating variable selection by a $\ell_1$ regularization. The microphone modes selection for improving the regression accuracy corresponds to the selection of the column vector in the regression coefficient matrix as shown in Fig.~\ref{fig:regressionimage}. Therefore, the group LASSO algorithm \citep{yuan2006model} that conducts the regularization using the group $\ell_1$ norm was employed. The regression coefficients matrix $\mathrm{\Psi}$ was obtained by optimizing following objective function:

\begin{equation}
\argmin_{\mathrm{\Psi}\in\mathbb{R}^{r_{\rm PIV}\times r_{\rm MIC}}}~~\frac{1}{2}\|\mathrm{\Psi}\mathbf{ Z}_\mathrm{MIC}-\mathbf{Z}_\mathrm{PIV}\|_2^2+\lambda \sum_{i=1}^{r_{\mathrm{MIC}}}\|\mathbf{\psi}_i\|_2,
\end{equation}
where the notation $\|\circ\|_2$ indicates the $\ell_2$ norm, and $\mathbf{\psi}_i$ is the $i$-th column vector of $\mathrm{\Psi}$. This objective function was optimized by the fast iterative shrinkage thresholding algorithm (FISTA) \citep{beck2009fast}. Here, the regularization parameter $\lambda$ that controls the sparsity of the regression coefficients matrix $\mathrm{\Psi}$ is one of the hyperparameters of the proposed method. When the regularization parameter $\lambda$ is large, the number of the selected microphone modes becomes less, and the regression coefficient matrix $\mathrm{\Psi}$ becomes sparse in the row direction.

Here, $\mathrm{\Phi}$ is defined as the matrix that consists of the non-zero components of the regression coefficient matrix $\mathrm{\Psi}$, as illustrated in Fig.~\ref{fig:regressionimage}. The matrix $\mathrm{\Phi}$ is interpreted as the space of PIV mode that can be reconstructed by the microphone modes. Therefore, the PIV mode coefficients that can be reconstructed by the microphone modes $\mathbf{Z}_\mathrm{\Phi}$ is defined as follows:

\begin{equation}
\mathbf{Z}_\mathrm{\Phi}=\mathrm{\Phi} \mathrm{\Phi}^{\dag} \mathbf{Z}_\mathrm{PIV},
\label{eqn:pivprojection}
\end{equation}

\noindent
where $\circ^\dag$ indicates the pseudo-inverse operation. The matrix $\mathbf{Z}_\mathrm{\Phi}$ is a projection of the $\mathbf{Z}_\mathrm{PIV}$ onto the $\mathrm{\Phi}$ space, and the projection ratio $PR$ is defined using the Frobenius norm of the matrices:

\begin{equation}
PR = 100 \times\left(1-\sqrt{\frac{\left\|\mathbf{Z}_{\mathrm{PIV}}-\mathbf{Z}_{\mathrm{\Phi}}\right\|_{F}^{2}}{\left\|\mathbf{Z}_{\mathrm{PIV}}\right\|_{F}^{2}}}\right).
\label{eqn:projectionratio}
\end{equation}

\noindent
This projection ratio indicates the reproducibility of the PIV data that can be reconstructed by the microphone modes with respect to the original reduced-order PIV data. In other words, this ratio represents the upper limit of the total energy of the phenomena that can be reconstructed by the present superresolution measurement.

\subsection{Reconstruction of the time-resolved PIV modes}
The time-resolved coefficient of PIV modes can simply be estimated using the regression coefficient matrix and the time-resolved microphone modes as follows:

\begin{equation}
\hat{\mathbf{Z}}_\mathrm{\Phi}=\mathrm{\Psi} \mathbf{Z}_\mathrm{MIC},
\label{eqn:reconstruction}
\end{equation}

\noindent
where $\hat{\mathbf{Z}}_\mathrm{\Phi}$ is the estimated time-resolved PIV modes coefficients. Consequently, the time-resolved velocity field can be reconstructed using the spatial mode of PIV data:

\begin{equation}
\hat{\mathbf{X}}_\mathrm{PIV}= \mathbf{U}_\mathrm{PIV}^{(r_{\mathrm{PIV}})} \hat{\mathbf{Z}}_\mathrm{\Phi}.
\label{eqn:reconstructvelocity}
\end{equation}

\noindent
Here, to evaluate the performance of the linear regression model, the model reconstruction error $E$ is also defined in addition to the projection ratio:

\begin{equation}
E = 100 \times\sqrt{\frac{\left\|\mathbf{Z}_{\mathrm{\Phi}}-\hat{\mathbf{Z}}_{\mathrm{\Phi}}\right\|_{F}^{2}}{\left\|\mathbf{Z}_{\mathrm{\Phi}}\right\|_{F}^{2}}}.
\label{eqn:reconstructionerror}
\end{equation}

\noindent
Equation~\ref{eqn:reconstructionerror} only evaluates the model reconstruction error of the linear regression model in the space of $\mathbf{Z}_\mathrm{\Phi}$, and does not directly indicate the reproducibility of the original reduced-order PIV data. Therefore, the quantitative performance of the superresolution measurement was evaluated using both Eqs.~\ref{eqn:projectionratio} and \ref{eqn:reconstructionerror}.

\subsection{Randomized cross-validation}
\begin{figure}[H]
    \centering
    \includegraphics[scale=0.6]{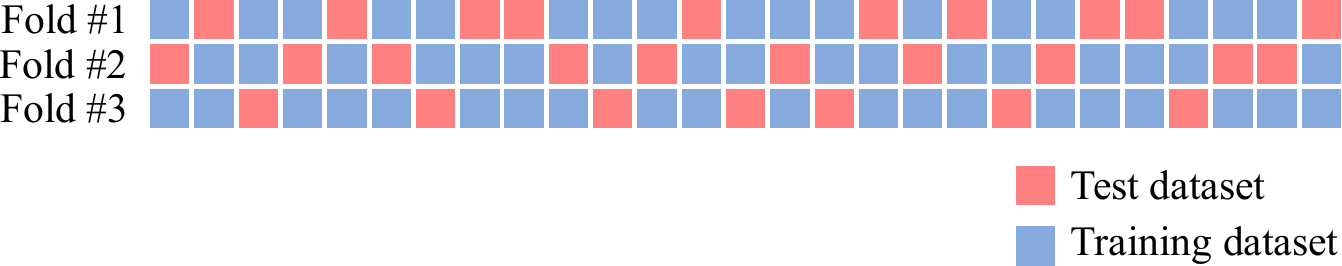}
    \caption{Schematic image of the randomized cross-validation.}
    \label{fig:randomizedCV}
\end{figure}

The present study evaluates the generalization performance of the superresolution measurement using a randomized $k$-fold cross-validation with $k=10$. Figure~\ref{fig:randomizedCV} depicts the schematic image of the randomized $k$-fold cross-validation while $k$ was set to be three for graphical explanation. The randomized cross-validation randomly selects the test and training data without overlapping, and sequential test and training datasets are not employed at all. This randomized cross-validation minimizes the effects of the trend components of the experimental data, whereas the trend component appears in the experimental data due to the slight changes in the experimental setup or the jet condition during the measurement. This procedure leads to the evaluation of the generalized performance of the superresolution measurement of non-trend components of target flow fields, which is of interest in the present study.

\section{Basic Characteristics of the Supersonic Jet}
\begin{figure}[H]
    \centering
    \includegraphics[scale=0.6]{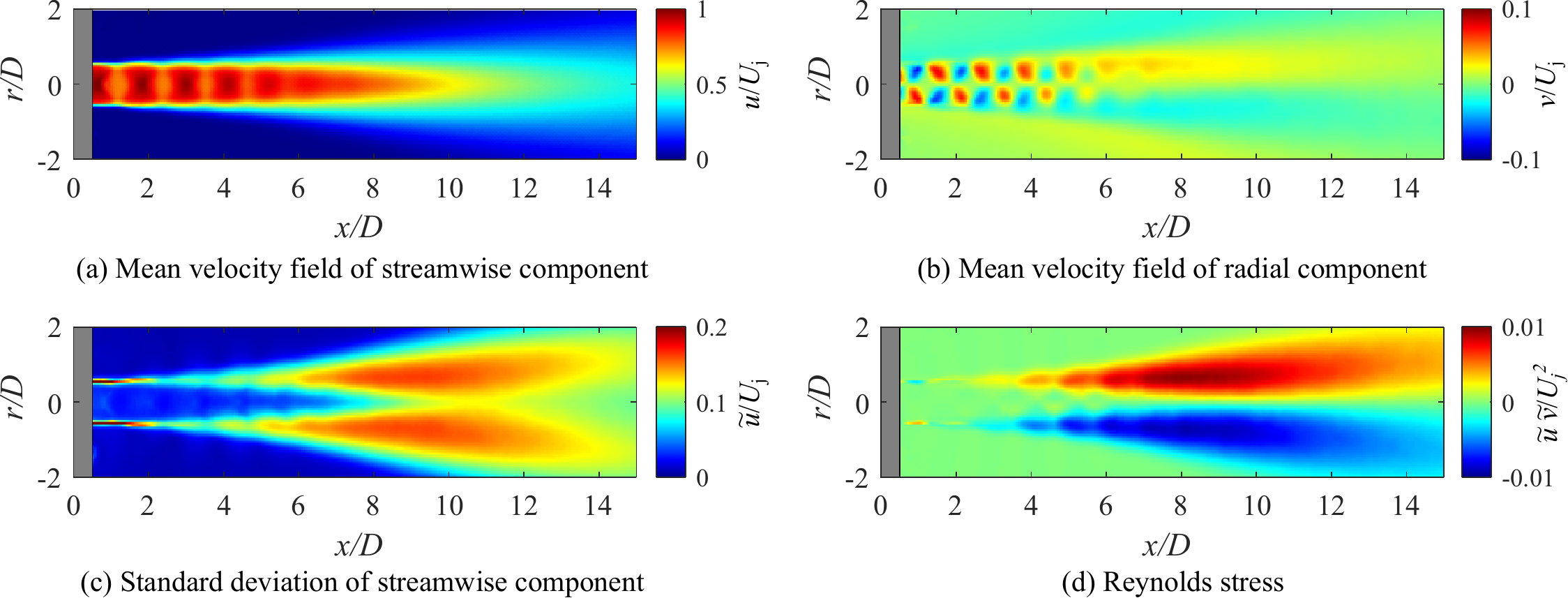}
    \caption{Basic characteristics of the velocity field.}
    \label{fig:RawVelocity}
\end{figure}

Figure~\ref{fig:RawVelocity} shows the mean velocity fields of the streamwise and radial components, the standard deviation of the streamwise velocity, and the Reynolds stress, respectively. The velocities are nondimensionalized by the theoretical streamwise velocity at the nozzle exit $U_j=392$~m/s that is derived under the assumption of the isentropic flow. The mean velocity fields identify the shear layer development, the potential core, and the shock wave structures called shock cells. As the shear layer develops towards the downstream, the standard deviation and the Reynolds stress increases and reach the maximum values near the end of the potential core. The present study eliminates the region near the nozzle exit ($x/D\leq0.5$) because the thinner shear layer may induce an error of the PIV algorithm due to the lack of the spatial resolution. The maximum fluctuation and Reynolds stress are observed at $x/D\approx8$ which is near the end of the potential core. This agrees well with the previous findings on the supersonic jet \citep{tam1995supersonic} and indicates the existence of a strong noise source. The screech frequency $f_s$ can be estimated using Eq.~\ref{eqn:peak_freq} proposed by \citet{powell1992observations}:

\begin{equation}
\frac{n}{f_{s}}=\frac{n_{sh}L_{sh}(1+M_{c})}{u_c},
\label{eqn:peak_freq}
\end{equation}

\noindent
where $M_c$, $u_c$, and $L_{sh}$ are the convection Mach number, the convection velocity, and the shock cell length, respectively. Variables $n$ and $n_{sh}$ are integers that indicate the number of screech cycles and the length of the noise source position. The observation of \citet{gao2010multi} showed that the integers are $n=5$ and $n_{sh}=5$ in the case of a Mach 1.35 supersonic jet that exhibits the flapping mode. The present study calculated the shock cell length $L_{sh}$ as the mean distance between the streamwise positions where the maximum velocity gradient is observed in the potential core. The calculated shock cell length is $L_{sh}=11.3$ mm, and the estimated screech frequency is $f_s=12.5$ kHz when $u_c=0.7U_j$. This estimated screech frequency is verified with the microphone data.

\begin{figure}[H]
    \centering
    \includegraphics[scale=0.6]{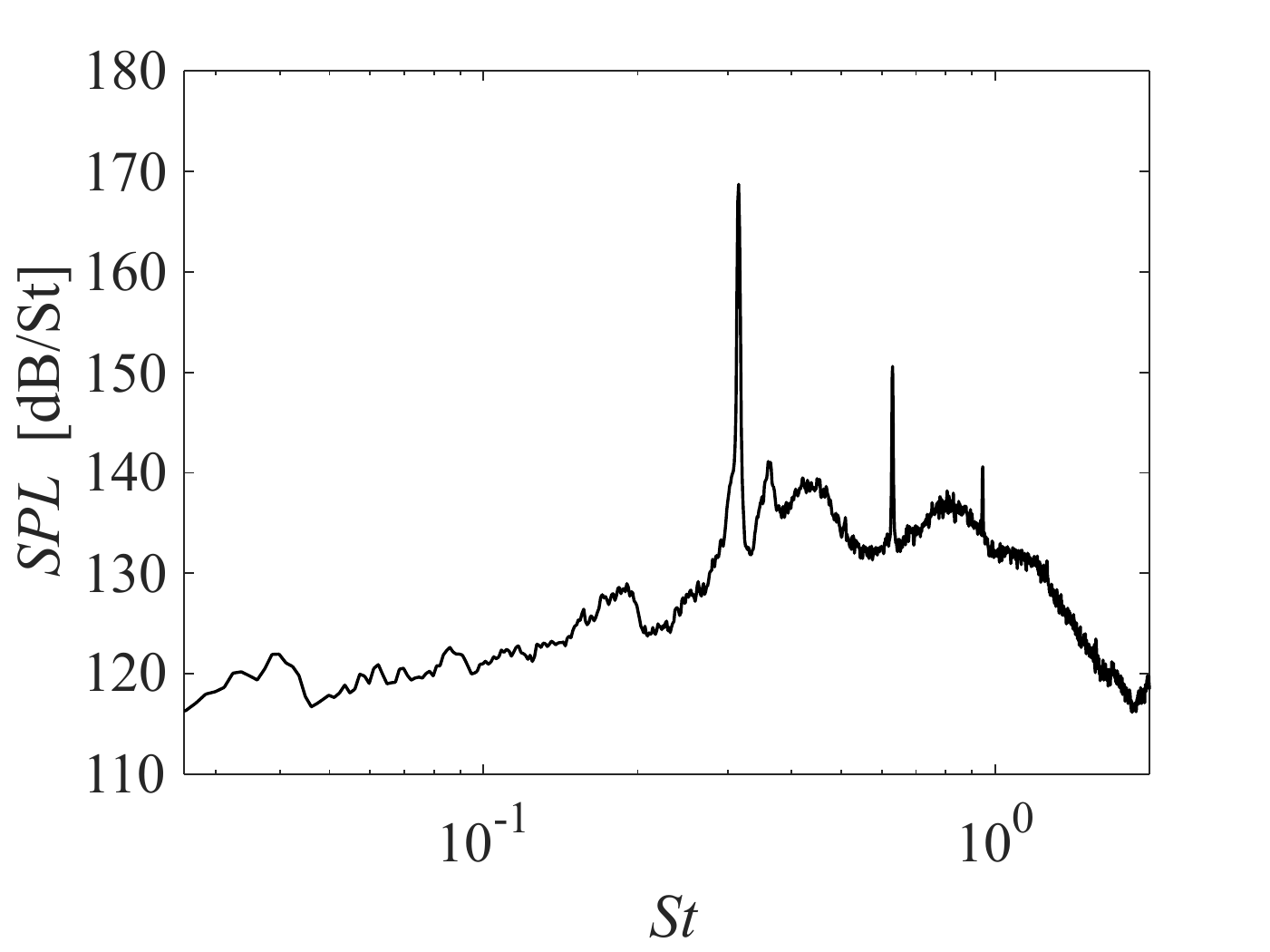}
    \caption{Acoustic spectra of the supersonic jet measured at ($x/D,\,r/D$) = ($0,\,4$).}
    \label{fig:SPL}
\end{figure}

Figure~\ref{fig:SPL} is the acoustic spectrum measured at ($x/D,\,r/D$) = ($0,\,4$). The Strouhal number was defined as follows:

\begin{equation}
St = \frac{fD}{U_j},
\label{eqn:strouhal}
\end{equation}

\noindent
where $f$ and $D$ are the frequency and the diameter at the nozzle exit, respectively. The spectrum agrees well with that of the report of \citet{andre2013broadband}. A distinct peak with over 20~dB amplitude is observed at 12.3~kHz ($St=0.32$). This is caused by the screech tone which is driven by a strong feedback loop and the resonance frequency agrees well with that estimated from Eq.~\ref{eqn:peak_freq}. The harmonics of the screech tone were also observed at 24.6 and 36.9~kHz. Therefore, the aeroacoustic field of the supersonic jet is dominated by the screech tone and the superresolution measurement in the present study mainly focuses on the reconstruction of the time-resolved velocity fluctuation associated with the screech tone generation.

\section{Results and discussion}
The superresolution measurement in the present study contains some hyperparameters in the analysis procedure as described in Sec~\ref{sec:CalculationProcedure}. Therefore, the effects of the hyperparameters on the model reconstruction error are firstly discussed, and then, the superresolved velocity fields are discussed in the case of the minimum model reconstruction error.

\subsection{Effect of dataset length $N$}
\begin{figure}[H]
    \centering
    \includegraphics[scale=0.6]{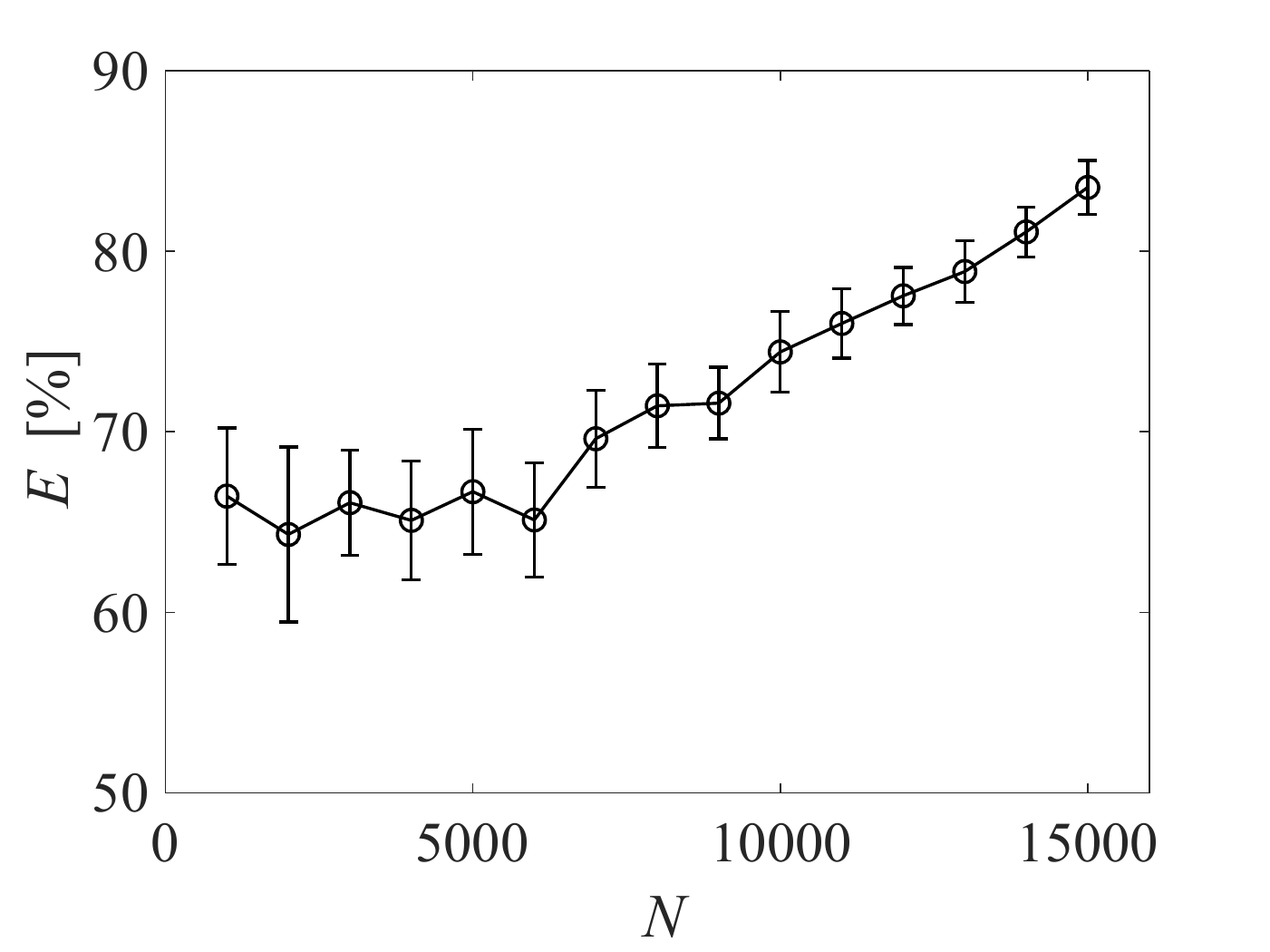}
    \caption{Effect of the dataset length $N$ on the model reconstruction error.}
    \label{fig:DataLengthEffect}
\end{figure}

Figure~\ref{fig:DataLengthEffect} shows the model reconstruction error with respect to the different dataset length $N$. In the cases of $N \leq 14,000$, the central part of the original dataset of $N=15,000$ is extracted by equally truncating both ends and the shorter datasets are reproduced. Therefore, the center data of the different datasets do not change with regardless of the dataset length, and the linear regression model is constructed based on the data which has almost the same trends in the dataset.

The minimum model reconstruction error was observed at $N=2,000$, and it does not significantly change until $N \leq 6,000$. The reconstruction accuracy may become worse due to insufficient training data when the dataset length is short ($N\leq1000$). On the other hand, the model reconstruction error monotonically increases at $N \ge 7,000$. This might be because the experimental setup or the jet condition may be slightly changed within the data acquisition duration when the dataset length is longer. Those may affect the POD mode coefficients as trend components of which the change is slower than that of the target phenomena, and cause an error in the construction of the regression model of the non-trend components in the present study. The reconstruction error can be reduced for a longer duration dataset if the good regression model including the trend components is constructed. However, the trend components of the present dataset could not be simply expressed in the linear regression model, though the present authors tried. Therefore, the present study employs the simple linear regression model and the dataset length of $N=2,000$, which minimizes the effect of the trend component, was selected.

\subsection{Effect of time-delay $n_{\mathrm{td}}$}
\label{sec:taueffect}
\begin{figure}[H]
    \centering
    \includegraphics[scale=0.6]{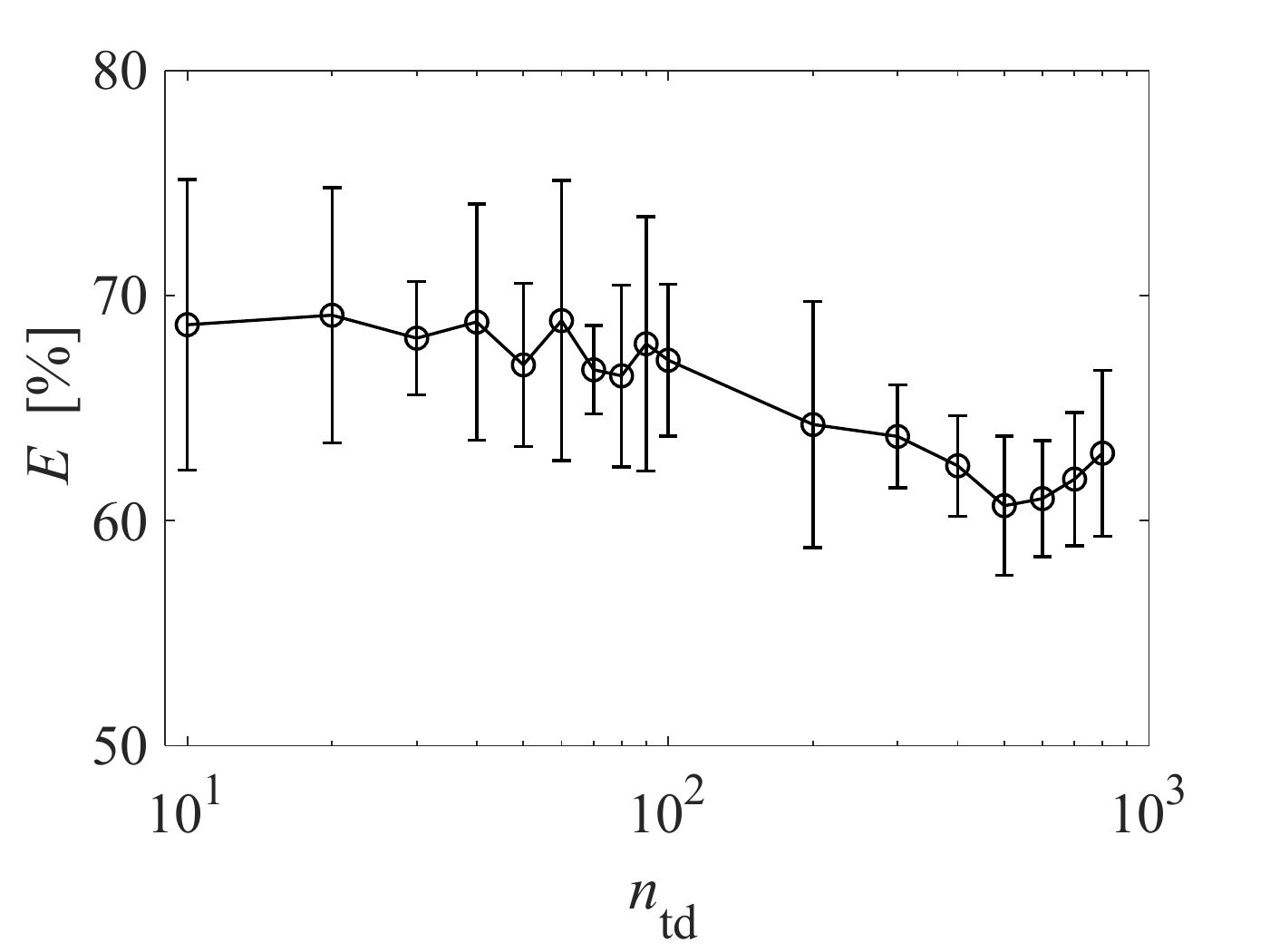}
    \caption{Effect of the time delay $n_{\mathrm{td}}$ on the model reconstruction error.}
    \label{fig:TimeDelayffect}
\end{figure}

Figure~\ref{fig:TimeDelayffect} shows the effect of the time delay $n_{\mathrm{td}}$ on the minimum model reconstruction error in the case of $N=2000$. The model reconstruction error decreases as the time delay $n_{\mathrm{td}}$ increases. The large $n_{\mathrm{td}}$ may strengthen the correlation between the PIV and microphone signals because the supersonic jet is fully turbulent and there is no exact time delay that is optimal for all snapshots. The minimum model reconstruction error is observed at $n_{\mathrm{td}}=500$, and it increases at $n_{\mathrm{td}} \ge 600$. The time-delay that achieves the minimum reconstruction error corresponds to the 2.5~ms.  Since the screech frequency was 12.3~kHz, the column vector of the microphone data matrix includes 61.5 periods of the fundamental screech frequency. The present study employed $n_{\mathrm{td}}=500$ for further analysis.

\subsection{Effect of regularization parameter $\lambda$}
\begin{figure}[H]
    \centering
    \includegraphics[scale=0.6]{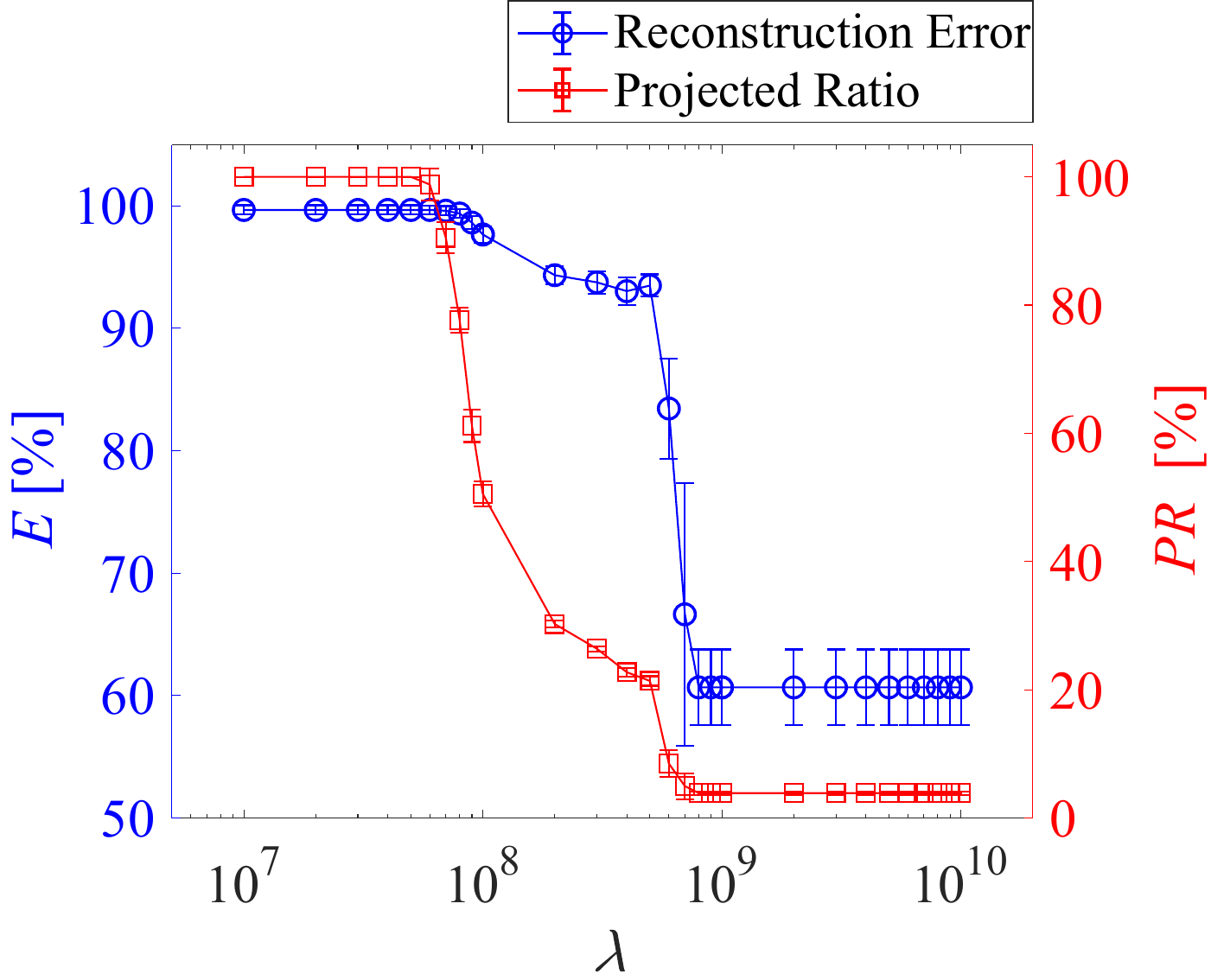}
    \caption{Relation of the model reconstruction error and the projected ratio.}
    \label{fig:RegularizationEffect_RE-PR}
\end{figure}

Figure~\ref{fig:RegularizationEffect_RE-PR} shows the effect of the regularization parameter $\lambda$ on the model reconstruction error and the projected ratio in the case of $N=2000$ for $n_{\mathrm{td}}=500$. As the regularization parameter increases, both model reconstruction error and projected ratio decrease. The large regularization parameter reduces the number of the selected microphone modes for the reconstruction. This means that the PIV space that can be reconstructed by the microphone modes becomes partial, resulting in the decrease in the projected ratio. Therefore, the phenomena that can be reconstructed by the linear regression model are limited when the regularization parameter is large. On the other hand, the decrease in the model reconstruction error indicates that the estimation accuracy of the linear regression model increases if the number of the selected microphone modes is small. Therefore, the linear regression model has a trade-off relationship between the reproducibility of the phenomena and the estimation accuracy.
The minimum model rec
onstruction error was 60.7 $\%$ observed at $\lambda \ge 10^9$. The selected microphone modes were the first four modes related to the screech tone, similar to those shown in Fig.~\ref{fig:POD-energy}. Since the velocity fluctuation associated with the screech tone is few compared to the entire fluid phenomena, the projected ratio of those cases was poor (3.88 $\%$). However, time-resolved velocity fluctuations related to the screech generation can be reconstructed. The results of the superresolved velocity field are discussed in the next section.

\subsection{Superresolved velocity fields}
\begin{figure}[H]
    \centering
    \includegraphics[scale=0.6]{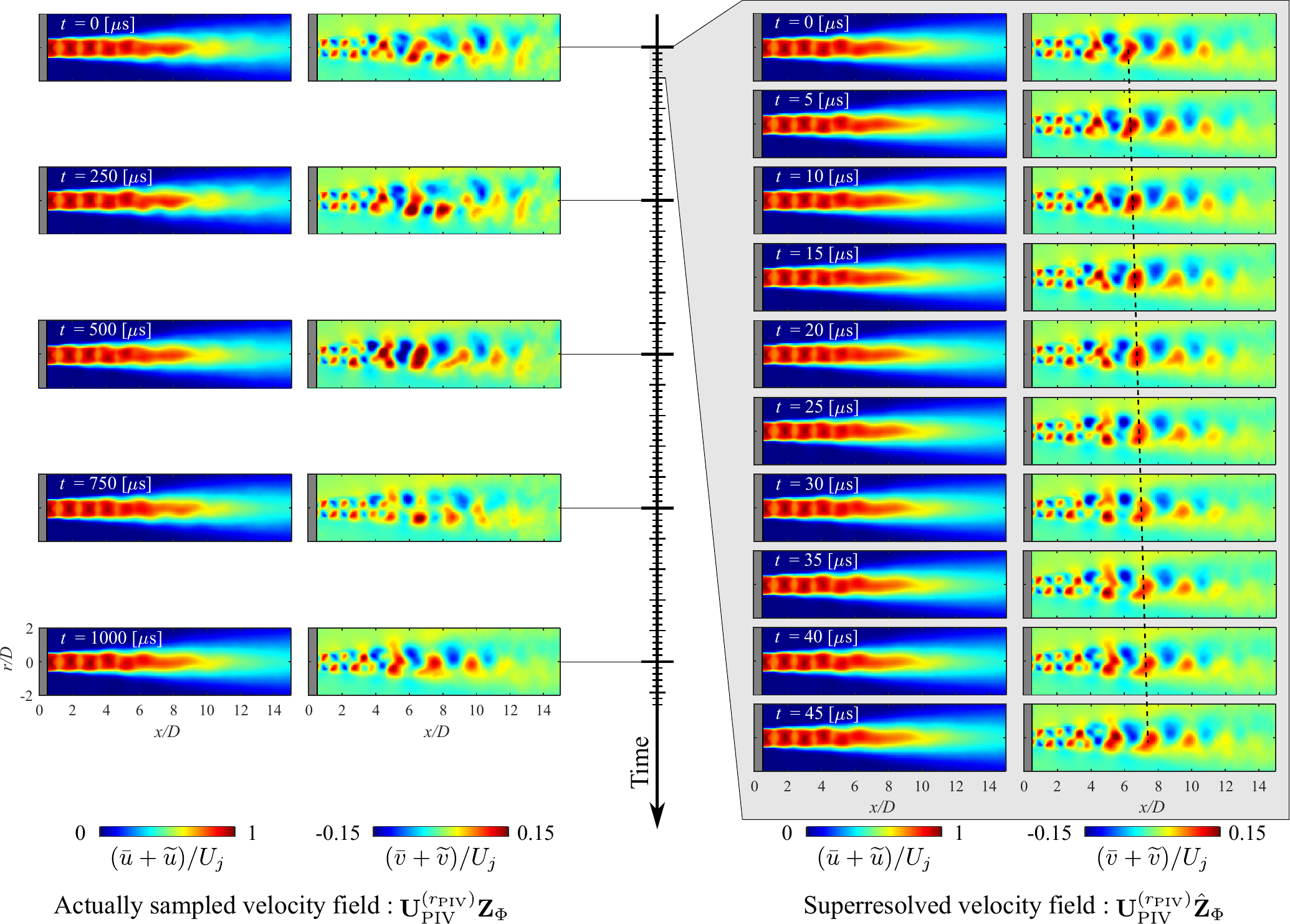}
    \caption{Snapshots of the actually sampled velocity field $\mathbf{U}_\mathrm{PIV}^{(r_{\mathrm{PIV}})} \mathbf{Z}_\mathrm{\Phi}$ and superresolved velocity fields $\mathbf{U}_\mathrm{PIV}^{(r_{\mathrm{PIV}})} \hat{\mathbf{Z}}_\mathrm{\Phi}$. See also the movie in the supplemental materials.}
    \label{fig:SRsnapshots}
\end{figure}

The superresolution measurement was performed using the parameters that can minimize the model reconstruction error as discussed in the previous sections. The dataset length $N$, the time-delay $n_{td}$, and the regularization parameter $\lambda$ were set to be 2,000, 500, and $10^9$, respectively. The group LASSO algorithm left only the first four microphone modes in the regression 
with tested parameter range. Figure~\ref{fig:SRsnapshots} shows the snapshots of the actually sampled and superresolved velocity fields. Note that the sampling rate of the superresolved snapshots is 200~kHz corresponding to that of the acoustic measurement. Here, actually sampled velocity field is calculated as $\mathbf{U}_\mathrm{PIV}^{(r_{\mathrm{PIV}})} \mathbf{Z}_\mathrm{\Phi}$, which is a projection of velocity field onto the space that the linear regression model can express. This is apart from the raw PIV data without the low-dimensionalization and the low-dimensionalized PIV data. The dashed line in this figure indicates the convection of the coherent structure observed in $\bar{v}+\widetilde{v}$. The superresoloved velocity field at $t=0$~$\mu$s qualitatively agrees with the actually sampled one taken at the same time while the actually sampled velocity field does not resolve the convection of the coherent structures due to the insufficient sampling rate. The movie that compares the superresolved and actually sampled velocity field is available in the supplementary material (Online Resource 1). The superresolved result shows the smooth convection of the flow structure while the actually sampled velocity field cannot illustrate the time-resolved fluid motion. The velocity distribution ($\bar{u}+\widetilde{u}$) exhibits the flapping motion of the fourth and fifth shock cells in the potential core as shown in Fig.~\ref{fig:SRsnapshots}. This characteristic fluctuation qualitatively agrees well with the characteristics of the screech mode B (flapping mode) \citep{powell1992observations,li2008numerical,li2020acoustic}. Although the unsteady fluctuation associated with the screech tone is observed, the convection of the large-scale structures at the downstream side was not observed in the proposed method. This implies that the phenomena that can be reconstructed from the microphone data are limited. In other words, the group LASSO algorithm only left PIV modes associated with the screech tone, and constructed the linear regression model.

\begin{figure}[H]
    \centering
    \includegraphics[scale=0.8]{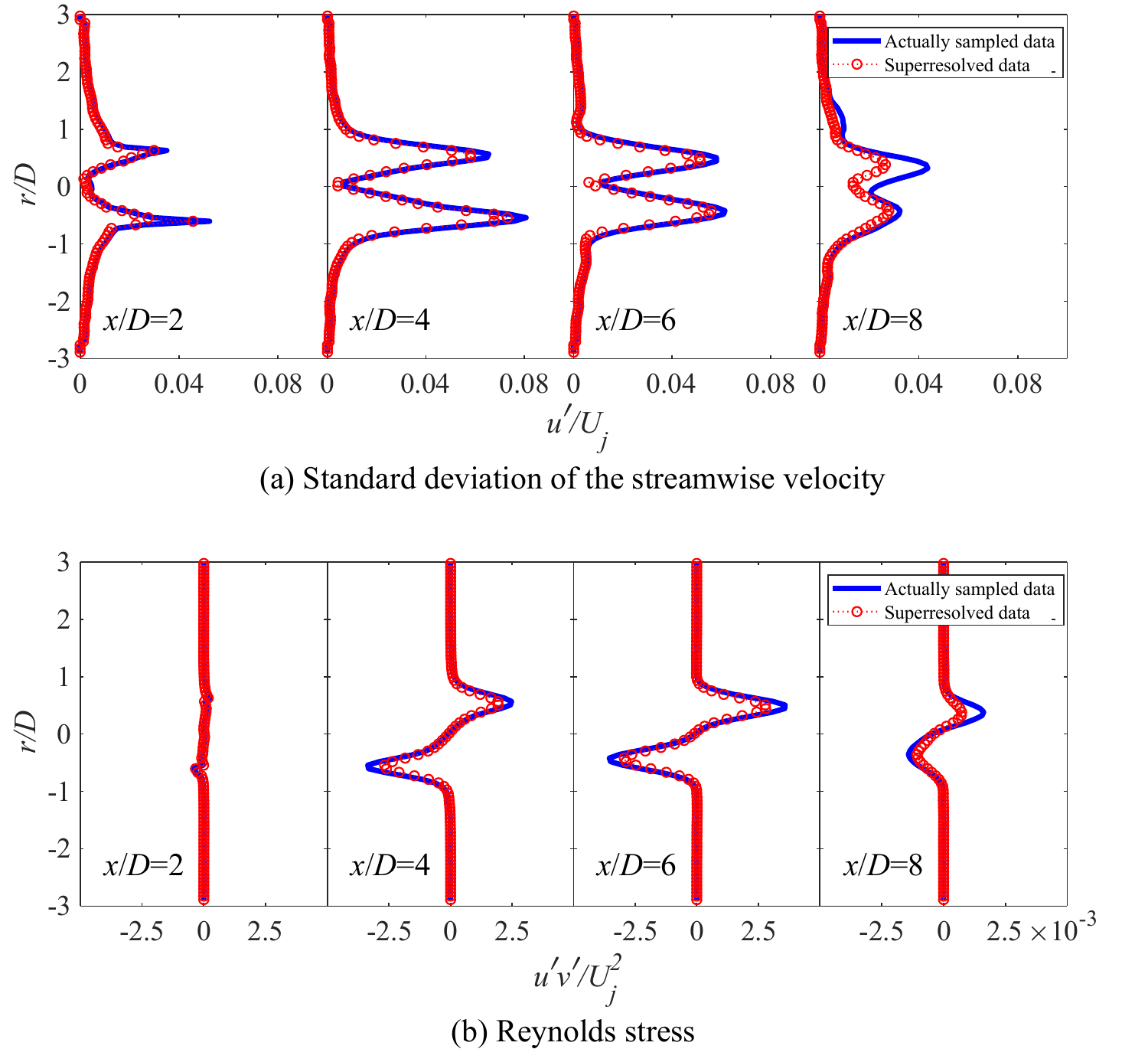}
    \caption{Radial distributions of the streamwise velocity fluctuation and the Reynolds stress.}
    \label{fig:SRvsAct_Ustd_Reys}
\end{figure}

The quantitative evaluation between actually sampled and superresolved velocity fields are provided as the velocity profiles shown in Fig.~\ref{fig:SRvsAct_Ustd_Reys}. Since the superresolved velocity fields can only be validated at the timing where the actually sampled velocity fields are available, the statistic quantities are compared. The standard deviation of the streamwise velocity and the Reynolds stress in Fig.~\ref{fig:SRvsAct_Ustd_Reys} are calculated using velocity fields for 0.05 seconds corresponding to 200 and 10,000 snapshots for actually sampled and superresolved velocity fields, respectively. The standard deviation and the Reynolds stress basically exhibit a similar profile between the actually sampled and superresolved data although the superresolved data underestimate the velocity fluctuations. This might be due to the measurement noise of the velocity field. Note that even the actually sampled velocity field underestimate the velocity fluctuations because it is a projection of velocity field onto the space that the linear regression model can express.

\begin{figure}[H]
    \centering
    \includegraphics[scale=0.6]{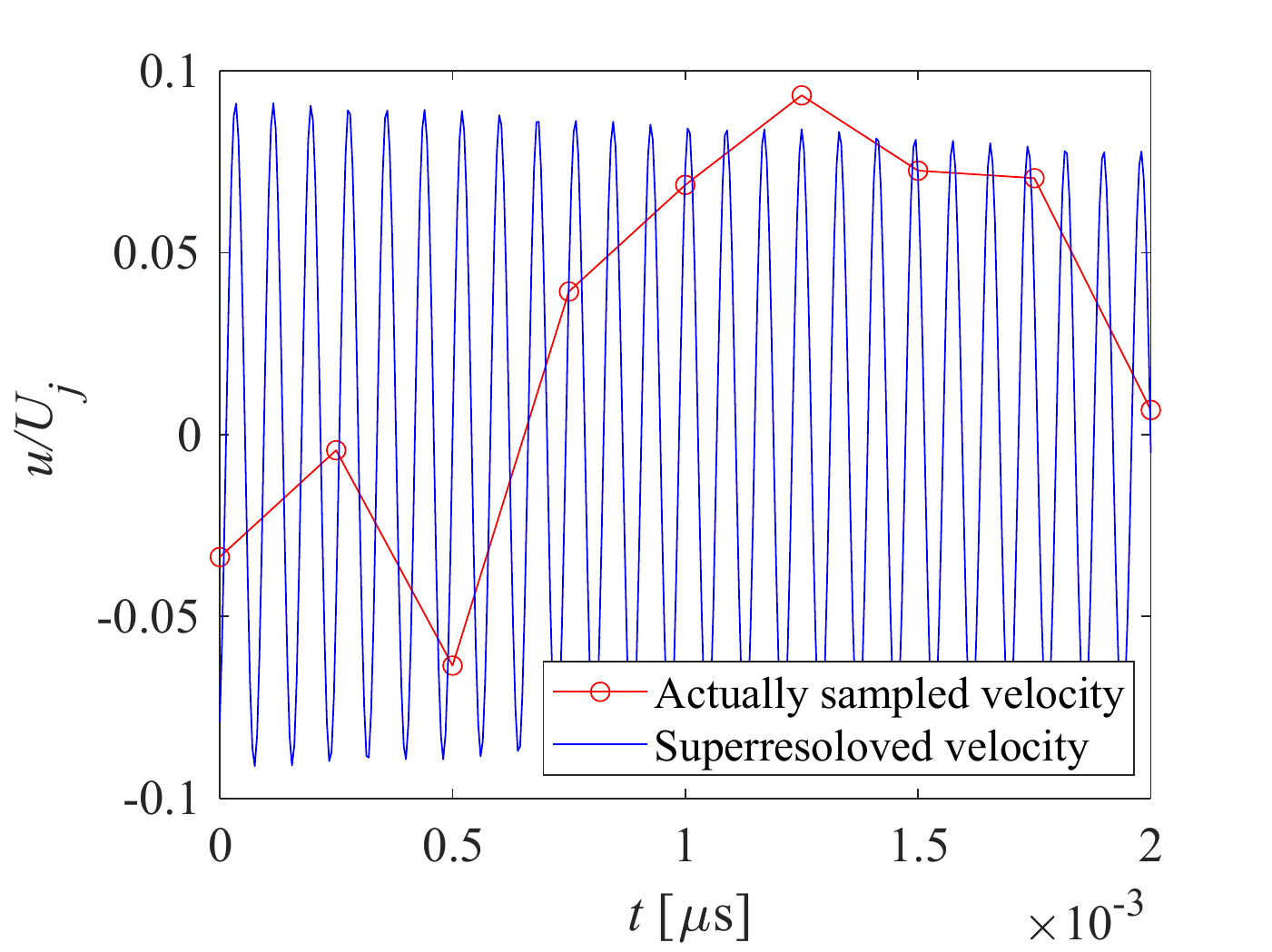}
    \caption{Comparison of the streamwise velocity at ($x/D$, $r/D$)=(4, 0.5).}
    \label{fig:SRvsAct_Velo}
\end{figure}

The streamwise velocity at ($x/D$, $r/D$)=(4, 0.5) was compared between the actually sampled and superresolved data as shown in Fig.~\ref{fig:SRvsAct_Velo}. The superresolved velocity sinusoidally oscillates over time while the actually sampled velocity does not resolve the unsteady fluctuations. Note that even the actually sampled velocity field underestimates the velocity fluctuations because it is a projection of the velocity field onto the space that the linear regression model can express, resulting in truncation of the original fluctuations.

\begin{figure}[H]
    \centering
    \includegraphics[scale=0.5]{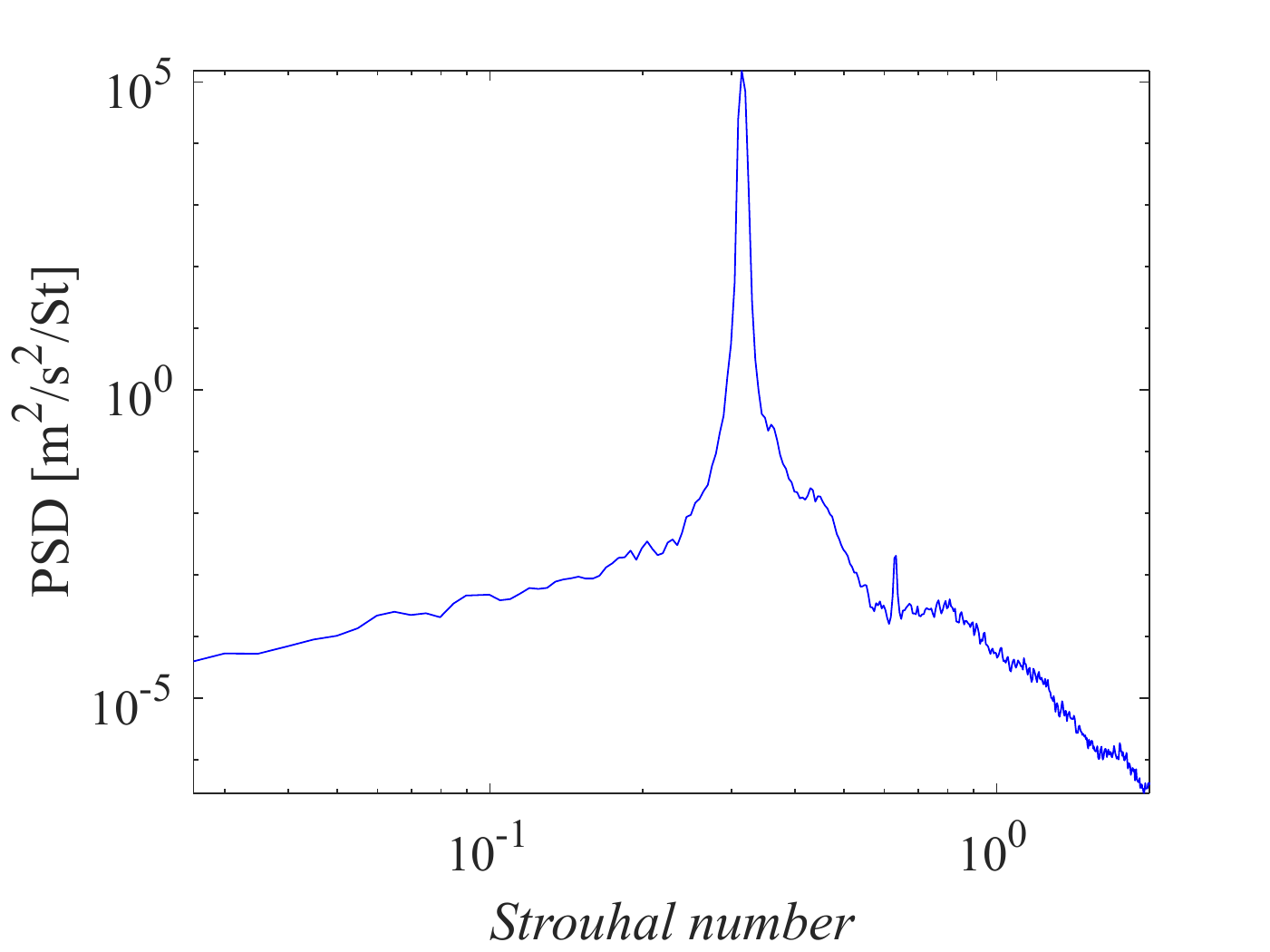}
    \caption{PSD of the superesoloved streamwise velocity at ($x/D$, $r/D$)=(4, 0.5).}
    \label{fig:PSDvel}
\end{figure}

To clarify the frequency characteristics of the superresoloved velocity fields, the power spectral density (PSD) was calculated using the superresolved streamwise velocity at the same position, as shown in Fig.~\ref{fig:PSDvel}. Figure~\ref{fig:PSDvel} shows a distinct peak at $St=0.31$, which is the same as that of the screech tone observed in the acoustic spectrum of Fig.~\ref{fig:SPL}. Moreover, the second harmonic of the screech tone is also observed in PSD. Therefore, the proposed method can reconstruct unsteady fluctuations of multiple frequency phenomena. Although conditional sampling or phase averaging has been widely used for the analysis of high-speed phenomena, those methods capture only the single frequency phenomena and assume the constant amplitude of the target phenomena within the measuring time. On the other hand, the proposed method can reconstruct the multiple frequency phenomena because this method uses the time-resolved microphone mode coefficients that include multiple frequency phenomena. This feature is one of the advantages of the proposed method.

\begin{figure}[H]
    \centering
    \includegraphics[scale=0.5]{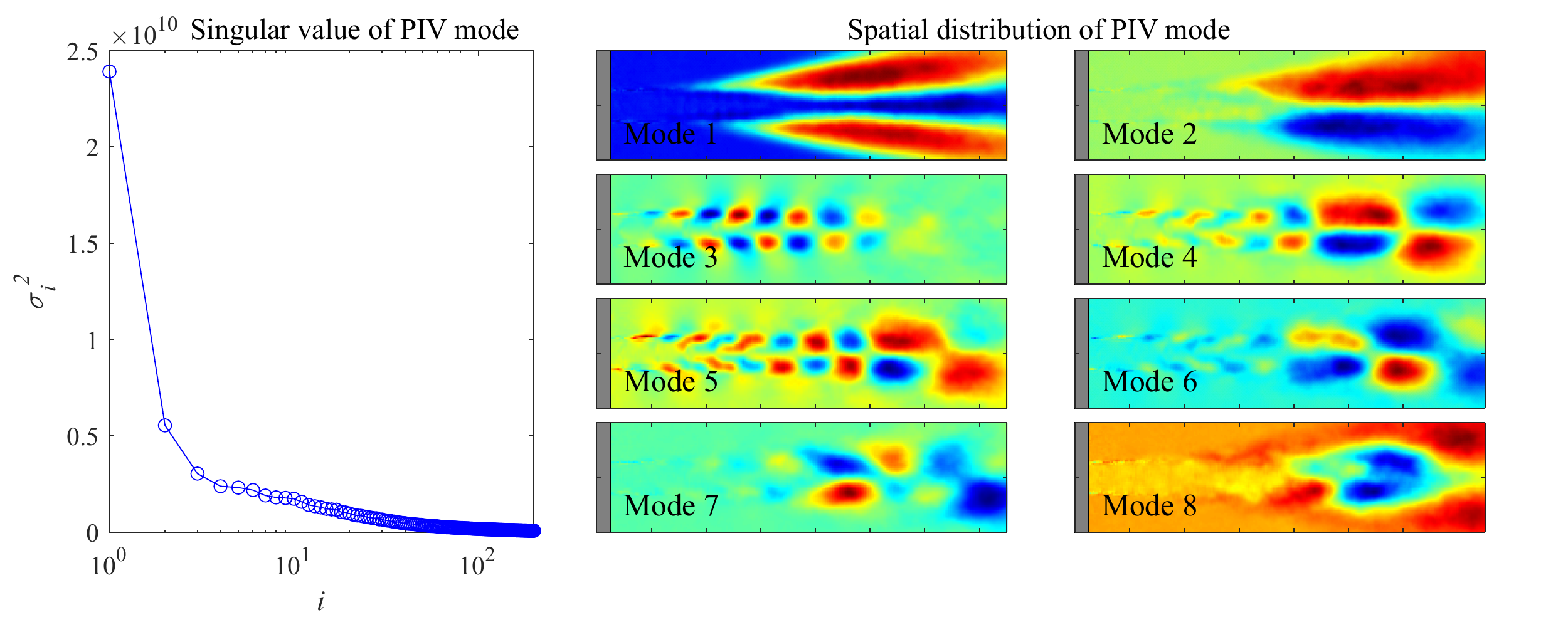}
    \caption{Characteristics of the PIV modes.}
    \label{fig:PIVmodes}
\end{figure}

It is worthwhile to discuss the applicability and limitation of the proposed method based on the spatial and temporal characteristics of each PIV mode in the superresolution result, since the superresolved velocity field is expressed as the superimposition of PIV modes. Figure~\ref{fig:PIVmodes} illustrates the singular value $\mathbf{S}_\mathrm{PIV}$ and the spatial distribution of streamwise velocity component in $\mathbf{U}_\mathrm{PIV}$. Here, this figure depicts the spatial distributions of only the first eight PIV modes. The first two PIV modes have relatively large singular values and exhibit large-scale fluctuations in the downstream shear layer. The third--seventh modes are asymmetric with respect to the jet axis and exhibit the cell structures in the shear layer region. These cell structures seem to express the velocity fluctuation caused by the shock cell oscillation that is a source of the screech tone.

\begin{figure}[H]
    \centering
    \includegraphics[scale=0.5]{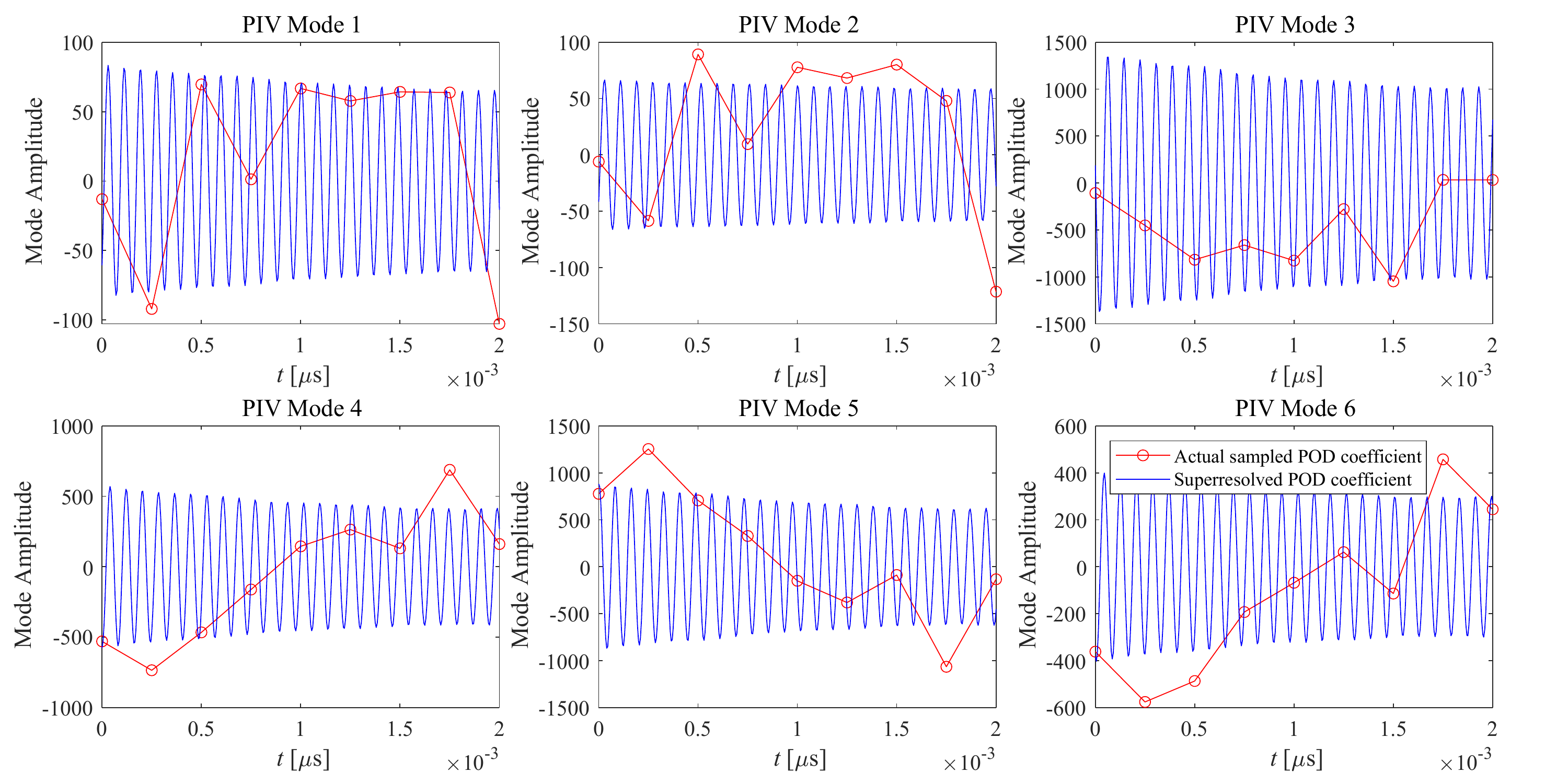}
    \caption{Time history of the first 6 PIV modes coefficients.}
    \label{fig:TimeHistoryPOD}
\end{figure}

\begin{figure}[H]
    \centering
    \includegraphics[scale=0.5]{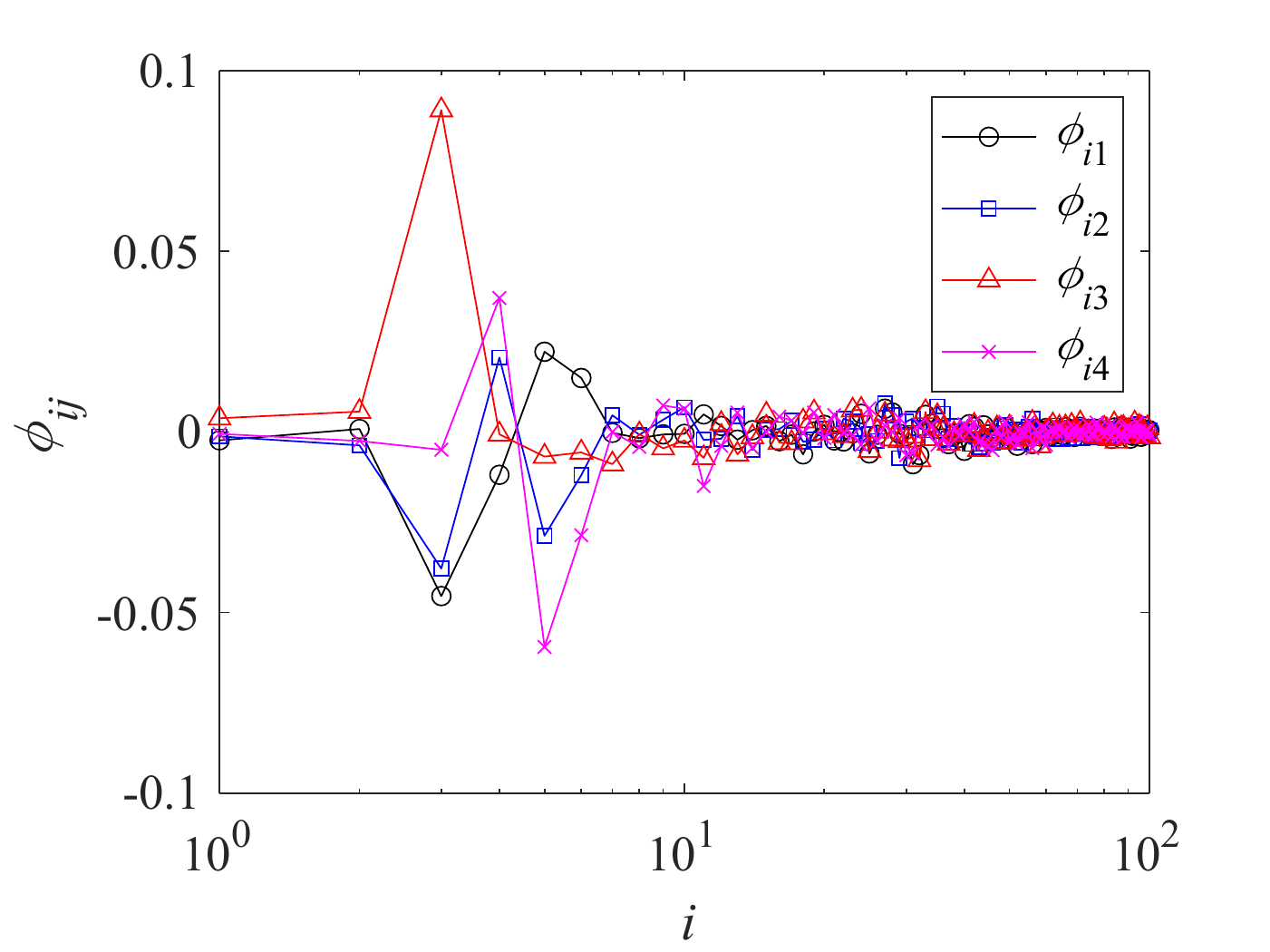}
    \caption{Regression coefficients in the matrix $\mathrm{\Phi}$.}
    \label{fig:RegressionCoef}
\end{figure}

Figure~\ref{fig:TimeHistoryPOD} shows the time histories of the mode coefficients of the first six PIV modes. This figure compares the superresolved POD coefficients with the actually sampled ones. Although the actually sampled POD coefficients do not obviously resolve the unsteady fluctuations, the sinusoidally oscillating mode coefficients are superresoloved. This frequency is expected to be the same as that of the screech tone. Here, the fluctuation amplitudes of the superresolved third--sixth PIV modes are much larger than those of the first two modes. The modes with large amplitude correspond to those that express the shock cell oscillation shown in Fig.~\ref{fig:PIVmodes}. This indicates that the proposed method selectively left the PIV modes that highly correlate with the screech tone. This is also observed in the entries of the regression coefficient matrix $\mathrm{\Phi}$ shown in Fig.~\ref{fig:RegressionCoef}. $\phi_{ij}$ is the entries of $\mathrm{\Phi}$ where $i$ and $j$ are the indices of the row and the column, respectively. In the present study, the numbers of the rows and columns of $\mathrm{\Phi}$ correspond to the numbers of the PIV and microphone modes, respectively. Most of the regression coefficients are close to zero except for $i$=3--6. Therefore, the LASSO regression left the third--sixth PIV modes that highly correlate with the selected microphone modes.

\begin{figure}[H]
    \centering
    \includegraphics[scale=0.5]{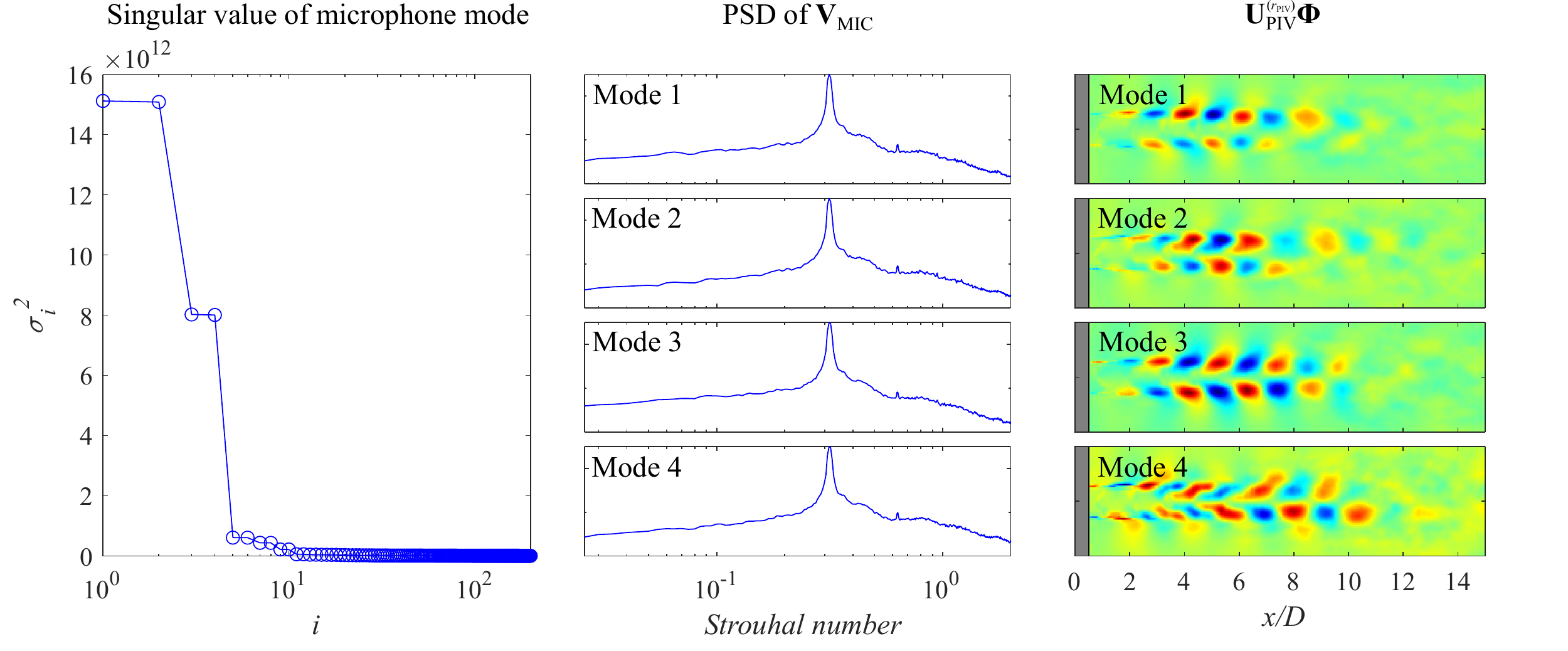}
    \caption{Characteristics of the selected microphone modes for the superresolution by the LASSO algorithm.}
    \label{fig:MICmodes}
\end{figure}

The characteristics of the selected microphone modes are summarized in Fig.~\ref{fig:MICmodes}. The singular values of the first four microphone modes are significantly larger than those of the other modes. Therefore, the dominant microphone modes are the first four modes, and thus, the LASSO regression left these microphone modes. The frequency characteristics of these microphone modes are shown as PSD in Fig.~\ref{fig:MICmodes}. PSDs indicate the distinct peak of the screech tone at $St=0.31$ and its second harmonic is also observed. These frequency characteristics agree well with those observed in the superresoloved velocity field shown in Fig.~\ref{fig:PSDvel}. Here, the velocity field reconstructed from the microphone modes can be calculated as the product of the spatial modes of PIV data $ \mathbf{U}_{\mathrm{PIV}}^{(r_{\mathrm{PIV}})} $ and the regression coefficient matrix $\mathrm{\Phi}$. The spatial distributions of $\mathbf{U}_{\mathrm{PIV}}^{(r_{\mathrm{PIV}})}\mathrm{\Phi}$ indicate the coherent structures in the shear layer region where the flapping mode is observed. Therefore, the superresolution measurement can reconstruct the phenomena that can be measured by the microphone.

\begin{figure}[H]
    \centering
    \includegraphics[scale=0.5]{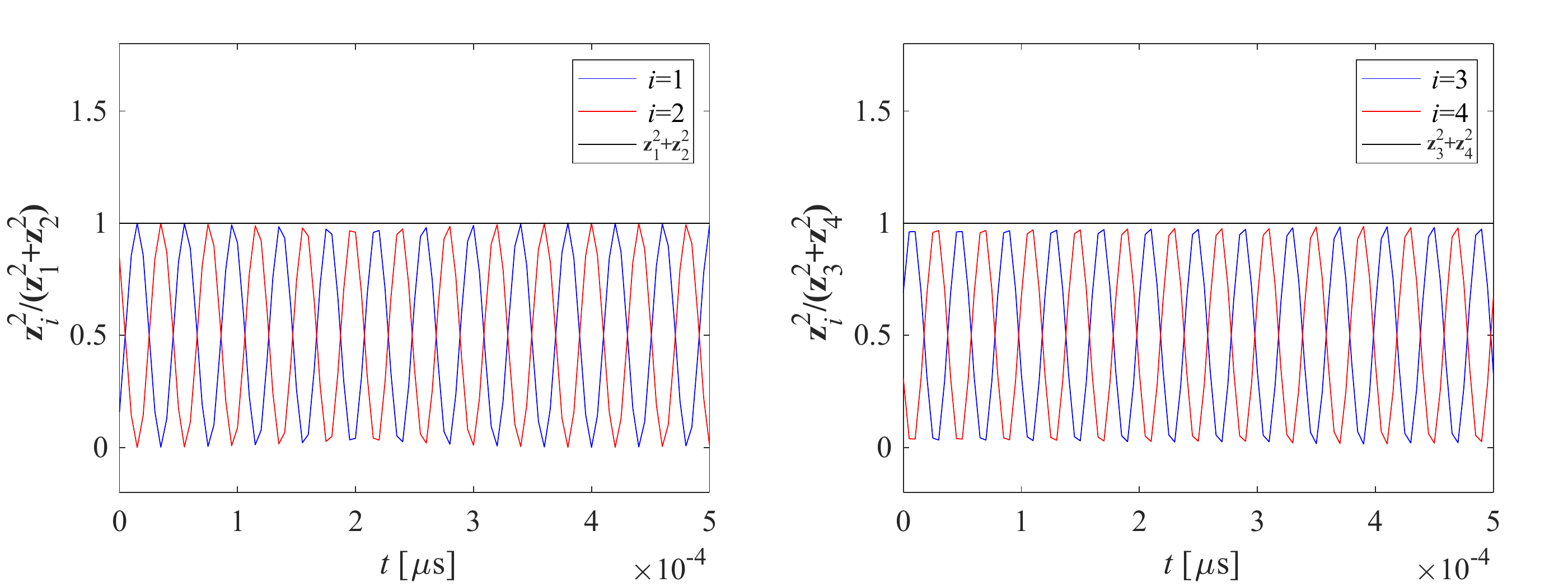}
    \caption{Snapshots of the actually sampled and superresolved velocity fields.}
    \label{fig:MicModeCoef}
\end{figure}

A relation of the temporal coefficients of the microphone modes was investigated because the superresolved velocity fields are reconstructed based on the time-resolved microphone data. Figure~\ref{fig:MicModeCoef} shows diagrams of the temporal coefficients of the microphone modes $\mathbf{z}_{i}$ where $\mathbf{z}_{i}$ is the $i$th row vector of $\mathbf{Z}_{\mathrm{MIC}}$. Since the first two pairs of microphone modes exhibit high energy compared to the others as shown in Fig.~\ref{fig:MICmodes}, the diagram is plotted for each paired mode. Each diagram shows a sinusoidal wave of the temporal coefficients, and the sum of corresponding coefficients is constant over time. Therefore, those microphone modes are paired and express the screech tone.

\section{Conclusions}
The present study proposed the framework for the spatiotemporal superresolution measurement using the non-time-resolved PIV and time-resolved acoustic measurements. The proposed framework is based on the sparse regression with the dimensionality reduction based on POD and was applied to a Mach 1.35 supersonic jet operated at the underexpanded condition. The PIV and acoustic measurements were simultaneously performed with the sampling rates of 4~kHz and 200~kHz, respectively. POD is applied to PIV and microphone data matrices and the sparse linear regression model of the reduced-order data was calculated using the LASSO regression. The proposed framework contains the hyperparameters: the dataset length $N$, the time-delay $n_{td}$, and the regularization parameter $\lambda$. The effects of these hyperparameters were quantitatively evaluated through randomized cross-validation, and the parameters with which the minimum model reconstruction error can be achieved were $N=2,000$, $n_{td}=500$, and $\lambda=10^9$, respectively, in the present dataset. The obtained minimum error is much smaller than that in the previous study using the leading POD modes and the linear least-square regression.

The superresolved velocity field reconstructed with the parameters above illustrates the smooth convection of the velocity fluctuations associated with the screech tone, although the convection of the large-scale structures at the downstream side was not observed. PSD of the superresolved velocity showed distinct peaks at the first and second harmonics of the screech tone. These frequency characteristics agree well with those of the selected microphone modes by the LASSO algorithm. Therefore, the proposed framework can reconstruct the unsteady fluctuation with multiple frequency phenomena, although the reconstruction is limited to the phenomena that can be measured by the microphone. This feature is the advantage of the proposed method because the conventional conditional sampling or phase averaging are hard to reconstruct the multiple frequency phenomena.

\section*{Acknowledgements}
This work was partially supported by JSPS KAKENHI Grant Number JP18H03809, JP19KK0361, and JP20H00278, and the research grants from Shimadzu Science Foundation. T.~Nagata was supported by Japan Science and Technology Agency, CREST Grant Number JPMJCR1763.


\bibliography{xaerolab}

\begin{thebibliography}{60}
\providecommand{\natexlab}[1]{#1}
\providecommand{\url}[1]{{#1}}
\providecommand{\urlprefix}{URL }
\expandafter\ifx\csname urlstyle\endcsname\relax
  \providecommand{\doi}[1]{DOI~\discretionary{}{}{}#1}\else
  \providecommand{\doi}{DOI~\discretionary{}{}{}\begingroup
  \urlstyle{rm}\Url}\fi
\providecommand{\eprint}[2][]{\url{#2}}

\bibitem[{Andr{\'e} et~al.(2011)Andr{\'e}, Castelain, and
  Bailly}]{andre2011experimental}
Andr{\'e} B, Castelain T, Bailly C (2011) Experimental study of flight effects
  on screech in underexpanded jets. Physics of Fluids 23(12):126,102

\bibitem[{Andr{\'e} et~al.(2013)Andr{\'e}, Castelain, and
  Bailly}]{andre2013broadband}
Andr{\'e} B, Castelain T, Bailly C (2013) Broadband shock-associated noise in
  screeching and non-screeching underexpanded supersonic jets. AIAA journal
  51(3):665--673

\bibitem[{Arroyo et~al.(2019)Arroyo, Daviller, Puigt, Airiau, and
  Moreau}]{arroyo2019identification}
Arroyo CP, Daviller G, Puigt G, Airiau C, Moreau S (2019) Identification of
  temporal and spatial signatures of broadband shock-associated noise. Shock
  Waves 29(1):117--134

\bibitem[{Bailly and Fujii(2016)}]{bailly2016high}
Bailly C, Fujii K (2016) High-speed jet noise. Mechanical Engineering Reviews
  3(1):15--00,496

\bibitem[{Beck and Teboulle(2009)}]{beck2009fast}
Beck A, Teboulle M (2009) A fast iterative shrinkage-thresholding algorithm for
  linear inverse problems. SIAM journal on imaging sciences 2(1):183--202

\bibitem[{Beresh et~al.(2015)Beresh, Kearney, Wagner, Guildenbecher, Henfling,
  Spillers, Pruett, Jiang, Slipchenko, Mance et~al.}]{beresh2015pulse}
Beresh S, Kearney S, Wagner J, Guildenbecher D, Henfling J, Spillers R, Pruett
  B, Jiang N, Slipchenko M, Mance J, et~al. (2015) Pulse-burst piv in a
  high-speed wind tunnel. Measurement Science and Technology 26(9):095,305

\bibitem[{Berkooz et~al.(1993)Berkooz, Holmes, and Lumley}]{berkooz1993proper}
Berkooz G, Holmes P, Lumley JL (1993) The proper orthogonal decomposition in
  the analysis of turbulent flows. Annual review of fluid mechanics
  25(1):539--575, \doi{10.1146/annurev.fl.25.010193.002543}

\bibitem[{Brunton and Kutz(2019)}]{brunton2019data}
Brunton SL, Kutz JN (2019) Data-driven science and engineering: Machine
  learning, dynamical systems, and control. Cambridge University Press

\bibitem[{Durgesh and Naughton(2010)}]{durgesh2010multi}
Durgesh V, Naughton J (2010) Multi-time-delay lse-pod complementary approach
  applied to unsteady high-reynolds-number near wake flow. Experiments in
  fluids 49(3):571--583

\bibitem[{Edgington-Mitchell et~al.(2014)Edgington-Mitchell, Oberleithner,
  Honnery, and Soria}]{edgington2014coherent}
Edgington-Mitchell D, Oberleithner K, Honnery DR, Soria J (2014) Coherent
  structure and sound production in the helical mode of a screeching
  axisymmetric jet. Journal of Fluid Mechanics 748:822--847

\bibitem[{Edgington-Mitchell et~al.(2018)Edgington-Mitchell, Jaunet, Jordan,
  Towne, Soria, and Honnery}]{edgington2018upstream}
Edgington-Mitchell D, Jaunet V, Jordan P, Towne A, Soria J, Honnery D (2018)
  Upstream-travelling acoustic jet modes as a closure mechanism for screech.
  Journal of Fluid Mechanics 855

\bibitem[{Gao and Li(2010)}]{gao2010multi}
Gao J, Li X (2010) A multi-mode screech frequency prediction formula for
  circular supersonic jets. The Journal of the Acoustical Society of America
  127(3):1251--1257

\bibitem[{Gojon and Bogey(2017)}]{gojon2017numerical}
Gojon R, Bogey C (2017) Numerical study of the flow and the near acoustic
  fields of an underexpanded round free jet generating two screech tones.
  International Journal of Aeroacoustics 16(7-8):603--625

\bibitem[{K{\"a}hler et~al.(2002)K{\"a}hler, Sammler, and
  Kompenhans}]{kahler2002generation}
K{\"a}hler C, Sammler B, Kompenhans J (2002) Generation and control of tracer
  particles for optical flow investigations in air. Experiments in fluids
  33(6):736--742

\bibitem[{Kanda et~al.(to appear)Kanda, Nakai, Saito, Nonomura, and
  Asai}]{kanda2021feasibility}
Kanda N, Nakai K, Saito Y, Nonomura T, Asai K (to appear) Feasibility study on
  real-time observation of flow velocity field by sparse processing particle
  image velocimetry. TRANSACTIONS OF THE JAPAN SOCIETY FOR AERONAUTICAL AND
  SPACE SCIENCES

\bibitem[{Lee et~al.(2021)Lee, Ozawa, Haga, Nonomura, and
  Asai}]{lee2021comparison}
Lee C, Ozawa Y, Haga T, Nonomura T, Asai K (2021) Comparison of
  three-dimensional density distribution of numerical and experimental analysis
  for twin jets. Journal of Visualization 24(6):1173--1188

\bibitem[{Li and Ukeiley(2021)}]{li2021pressure}
Li S, Ukeiley L (2021) Pressure-informed velocity estimation in a subsonic jet.
  arXiv preprint arXiv:210607110

\bibitem[{Li and Gao(2008)}]{li2008numerical}
Li X, Gao J (2008) Numerical simulation of the three-dimensional screech
  phenomenon from a circular jet. Physics of Fluids 20(3):035,101

\bibitem[{Li et~al.(2021)Li, Liu, Hao, Zhang, and He}]{li2021screech}
Li X, Liu N, Hao P, Zhang X, He F (2021) Screech feedback loop and mode staging
  process of axisymmetric underexpanded jets. Experimental Thermal and Fluid
  Science 122:110,323

\bibitem[{Li et~al.(2020)Li, Zhang, Hao, and He}]{li2020acoustic}
Li XR, Zhang XW, Hao PF, He F (2020) Acoustic feedback loops for screech tones
  of underexpanded free round jets at different modes. Journal of Fluid
  Mechanics 902

\bibitem[{Lim et~al.(2020)Lim, Wei, Zang, Vevek, Mariani, New, and
  Cui}]{lim2020short}
Lim H, Wei X, Zang B, Vevek U, Mariani R, New T, Cui Y (2020) Short-time proper
  orthogonal decomposition of time-resolved schlieren images for transient jet
  screech characterization. Aerospace Science and Technology 107:106,276

\bibitem[{Lumley(1967)}]{lumley1967structure}
Lumley JL (1967) The structure of inhomogeneous turbulent flows. Atmospheric
  turbulence and radio wave propagation

\bibitem[{Mercier et~al.(2016)Mercier, Castelain, and
  Bailly}]{mercier2016schlieren}
Mercier B, Castelain T, Bailly C (2016) A schlieren and nearfield acoustic
  based experimental investigation of screech noise sources. In: 22nd AIAA/CEAS
  Aeroacoustics Conference, p 2799

\bibitem[{Mercier et~al.(2017)Mercier, Castelain, and
  Bailly}]{mercier2017experimental}
Mercier B, Castelain T, Bailly C (2017) Experimental characterisation of the
  screech feedback loop in underexpanded round jets. Journal of Fluid Mechanics
  824:202--229

\bibitem[{Nankai et~al.(2019)Nankai, Ozawa, Nonomura, and
  Asai}]{nankai2019linear}
Nankai K, Ozawa Y, Nonomura T, Asai K (2019) Linear reduced-order model based
  on piv data of flow field around airfoil. TRANSACTIONS OF THE JAPAN SOCIETY
  FOR AERONAUTICAL AND SPACE SCIENCES 62(4):227--235,
  \doi{10.2322/tjsass.62.227}

\bibitem[{Nickels et~al.(2020)Nickels, Ukeiley, Reger, and
  Cattafesta~III}]{nickels2020low}
Nickels A, Ukeiley L, Reger R, Cattafesta~III L (2020) Low-order estimation of
  the velocity, hydrodynamic pressure, and acoustic radiation for a
  three-dimensional turbulent wall jet. Experimental Thermal and Fluid Science
  116:110,101

\bibitem[{Nishikori(2022)}]{nishikori2022superresolution}
Nishikori H (2022) Superresolution measurement of supersonic jet (in
  {J}apanese). M.{S}. thesis, Tohoku University, Sendai, JAPAN

\bibitem[{Nonomura et~al.(2019)Nonomura, Nakano, Ozawa, Terakado, Yamamoto,
  Fujii, and Oyama}]{nonomura2019large}
Nonomura T, Nakano H, Ozawa Y, Terakado D, Yamamoto M, Fujii K, Oyama A (2019)
  Large eddy simulation of acoustic waves generated from a hot supersonic jet.
  Shock Waves pp 1--22

\bibitem[{Nonomura et~al.(2021{\natexlab{a}})Nonomura, Ibuki, Ozawa, Asai, and
  Oyama}]{nonomura2021generalized}
Nonomura T, Ibuki T, Ozawa Y, Asai K, Oyama A (2021{\natexlab{a}}) Generalized
  estimation methods of turbulent fluctuation of high-speed flow with
  single-pixel resolution particle image velocimetry. Measurement Science and
  Technology 32(12):125,306

\bibitem[{Nonomura et~al.(2021{\natexlab{b}})Nonomura, Nankai, Iwasaki, Komuro,
  and Asai}]{nonomura2021quantitative}
Nonomura T, Nankai K, Iwasaki Y, Komuro A, Asai K (2021{\natexlab{b}})
  Quantitative evaluation of predictability of linear reduced-order model based
  on particle-image-velocimetry data of separated flow field around airfoil.
  Experiments in Fluids 62:112

\bibitem[{Nonomura et~al.(2021{\natexlab{c}})Nonomura, Ozawa, Abe, and
  Fujii}]{nonomura2021computational}
Nonomura T, Ozawa Y, Abe Y, Fujii K (2021{\natexlab{c}}) Computational study on
  aeroacoustic fields of a transitional supersonic jet. The Journal of the
  Acoustical Society of America 149(6):4484--4502

\bibitem[{Ohmizu et~al.(2022)Ohmizu, Ozawa, Nagata, Nonomura, and
  Asai}]{ohmizu2022demonstration}
Ohmizu K, Ozawa Y, Nagata T, Nonomura T, Asai K (2022) Demonstration and
  verification of exact {DMD} analysis applied to double-pulsed schlieren image
  of supersonic impinging jet. Journal of visualization, to appear
  \doi{10.1007/s12650-022-00836-9}

\bibitem[{Ozawa et~al.(2018)Ozawa, Nonomura, Anyoji, Mamori, Fukushima, Oyama,
  Fujii, and Yamamoto}]{ozawa2018identification}
Ozawa Y, Nonomura T, Anyoji M, Mamori H, Fukushima N, Oyama A, Fujii K,
  Yamamoto M (2018) Identification of acoustic wave propagation pattern of a
  supersonic jet using frequency-domain pod. Transactions of the Japan Society
  for Aeronautical and Space Sciences 61(6):281--284

\bibitem[{Ozawa et~al.(2020{\natexlab{a}})Ozawa, Ibuki, Nonomura, Suzuki,
  Komuro, Ando, and Asai}]{ozawa2020single}
Ozawa Y, Ibuki T, Nonomura T, Suzuki K, Komuro A, Ando A, Asai K
  (2020{\natexlab{a}}) Single-pixel resolution velocity/convection velocity
  field of a supersonic jet measured by particle/schlieren image velocimetry.
  Experiments in Fluids 61(6)

\bibitem[{Ozawa et~al.(2020{\natexlab{b}})Ozawa, Nonomura, Oyama, and
  Asai}]{ozawa2020effect}
Ozawa Y, Nonomura T, Oyama A, Asai K (2020{\natexlab{b}}) Effect of the
  reynolds number on the aeroacoustic fields of a transitional supersonic jet.
  Physics of Fluids 32(4):046,108

\bibitem[{Ozawa et~al.(2021)Ozawa, Nagata, Nonomura, and Asai}]{ozawa2021pod}
Ozawa Y, Nagata T, Nonomura T, Asai K (2021) {POD}-based spatio-temporal
  superresolution measurement on a supersonic jet using {PIV} and near-field
  acoustic data. In: AIAA AVIATION 2021 FORUM, p 2106

\bibitem[{Panda(1999)}]{panda1999experimental}
Panda J (1999) An experimental investigation of screech noise generation.
  Journal of Fluid Mechanics 378:71--96

\bibitem[{Pineau and Bogey(2021{\natexlab{a}})}]{pineau2021links}
Pineau P, Bogey C (2021{\natexlab{a}}) Links between steepened mach waves and
  coherent structures for a supersonic jet. AIAA Journal 59(5):1673--1681

\bibitem[{Pineau and Bogey(2021{\natexlab{b}})}]{pineau2021numerical}
Pineau P, Bogey C (2021{\natexlab{b}}) Numerical investigation of wave
  steepening and shock coalescence near a cold mach 3 jet. The Journal of the
  Acoustical Society of America 149(1):357--370

\bibitem[{Powell(1953{\natexlab{a}})}]{powell1953noise}
Powell A (1953{\natexlab{a}}) The noise of choked jets. The Journal of the
  Acoustical Society of America 25(3):385--389

\bibitem[{Powell(1953{\natexlab{b}})}]{powell1953mechanism}
Powell A (1953{\natexlab{b}}) On the mechanism of choked jet noise. Proceedings
  of the Physical Society Section B 66(12):1039

\bibitem[{Powell et~al.(1992)Powell, Umeda, and Ishii}]{powell1992observations}
Powell A, Umeda Y, Ishii R (1992) Observations of the oscillation modes of
  choked circular jets. The Journal of the Acoustical Society of America
  92(5):2823--2836

\bibitem[{Price et~al.(2021)Price, Gragston, and Kreth}]{price2021supersonic}
Price TJ, Gragston M, Kreth PA (2021) Supersonic underexpanded jet features
  extracted from modal analyses of high-speed optical diagnostics. AIAA Journal
  pp 1--18

\bibitem[{Raman(1999)}]{raman1999supersonic}
Raman G (1999) Supersonic jet screech: half-century from powell to the present.
  Journal of Sound and Vibration 225(3):543--571

\bibitem[{Rao et~al.(2020)Rao, Kushari, and Chandra~Mandal}]{rao2020screech}
Rao AN, Kushari A, Chandra~Mandal A (2020) Screech characteristics of
  under-expanded high aspect ratio elliptic jet. Physics of Fluids
  32(7):076,106

\bibitem[{Scharnowski et~al.(2012)Scharnowski, Hain, and
  K{\"a}hler}]{scharnowski2012reynolds}
Scharnowski S, Hain R, K{\"a}hler CJ (2012) Reynolds stress estimation up to
  single-pixel resolution using piv-measurements. Experiments in fluids
  52(4):985--1002

\bibitem[{Schmid(2010)}]{schmid2010dynamic}
Schmid PJ (2010) Dynamic mode decomposition of numerical and experimental data.
  Journal of Fluid Mechanics 656(July 2010):5--28,
  \doi{10.1017/S0022112010001217}, \eprint{arXiv:1312.0041v1}

\bibitem[{Shariff and Manning(2013)}]{shariff2013ray}
Shariff K, Manning TA (2013) A ray tracing study of shock leakage in a model
  supersonic jet. Physics of Fluids 25(7):076,103

\bibitem[{Suzuki and Lele(2003)}]{suzuki2003shock}
Suzuki T, Lele SK (2003) Shock leakage through an unsteady vortex-laden mixing
  layer: application to jet screech. Journal of Fluid Mechanics 490:139--167

\bibitem[{Suzuki et~al.(2020)Suzuki, Chatellier, and David}]{suzuki2020few}
Suzuki T, Chatellier L, David L (2020) A few techniques to improve data-driven
  reduced-order simulations for unsteady flows. Computers \& Fluids 201:104,455

\bibitem[{Taira et~al.(2017)Taira, Brunton, Dawson, Rowley, Colonius, McKeon,
  Schmidt, Gordeyev, Theofilis, and Ukeiley}]{taira2017modal}
Taira K, Brunton SL, Dawson ST, Rowley CW, Colonius T, McKeon BJ, Schmidt OT,
  Gordeyev S, Theofilis V, Ukeiley LS (2017) Modal analysis of fluid flows: An
  overview. AIAA Journal pp 4013--4041, \doi{10.2514/1.J056060}

\bibitem[{Tam(1995)}]{tam1995supersonic}
Tam CK (1995) Supersonic jet noise. Annual review of fluid mechanics
  27(1):17--43

\bibitem[{Tan et~al.(2019)Tan, Honnery, Kalyan, Gryazev, Karabasov, and
  Edgington-Mitchell}]{tan2019correlation}
Tan DJ, Honnery D, Kalyan A, Gryazev V, Karabasov SA, Edgington-Mitchell D
  (2019) Correlation analysis of high-resolution particle image velocimetry
  data of screeching jets. AIAA journal 57(2):735--748

\bibitem[{Tibshirani(1996)}]{tibshirani1996regression}
Tibshirani R (1996) Regression shrinkage and selection via the lasso. Journal
  of the Royal Statistical Society: Series B (Methodological) 58(1):267--288

\bibitem[{Tinney et~al.(2008)Tinney, Ukeiley, and Glauser}]{tinney2008low}
Tinney C, Ukeiley L, Glauser MN (2008) Low-dimensional characteristics of a
  transonic jet. part 2. estimate and far-field prediction. Journal of Fluid
  Mechanics 615:53--92

\bibitem[{Tu et~al.(2013{\natexlab{a}})Tu, Griffin, Hart, Rowley, Cattafesta,
  and Ukeiley}]{tu2013integration}
Tu JH, Griffin J, Hart A, Rowley CW, Cattafesta LN, Ukeiley LS
  (2013{\natexlab{a}}) Integration of non-time-resolved piv and time-resolved
  velocity point sensors for dynamic estimation of velocity fields. Experiments
  in fluids 54(2):1--20

\bibitem[{Tu et~al.(2013{\natexlab{b}})Tu, Rowley, Luchtenburg, Brunton, and
  Kutz}]{tu2013dynamic}
Tu JH, Rowley CW, Luchtenburg DM, Brunton SL, Kutz JN (2013{\natexlab{b}}) On
  dynamic mode decomposition: theory and applications. arXiv preprint
  arXiv:13120041

\bibitem[{Westerweel et~al.(2004)Westerweel, Geelhoed, and
  Lindken}]{westerweel2004single}
Westerweel J, Geelhoed P, Lindken R (2004) Single-pixel resolution ensemble
  correlation for micro-piv applications. Experiments in Fluids 37(3):375--384

\bibitem[{Yuan and Lin(2006)}]{yuan2006model}
Yuan M, Lin Y (2006) Model selection and estimation in regression with grouped
  variables. Journal of the Royal Statistical Society: Series B (Statistical
  Methodology) 68(1):49--67

\bibitem[{Zhang et~al.(2020)Zhang, Cattafesta, and Ukeiley}]{zhang2020spectral}
Zhang Y, Cattafesta LN, Ukeiley L (2020) Spectral analysis modal methods
  (samms) using non-time-resolved piv. Experiments in Fluids 61(11):1--12

\end{thebibliography}
\bibliographystyle{spbasic}
\end{document}